\documentclass[preprint]{aastex}

\usepackage{amssymb}
\usepackage{natbib}
\usepackage{amsmath}
\usepackage{wasysym}
\usepackage{graphicx}
\usepackage{wrapfig}
\usepackage{multirow}
\usepackage{array}
\usepackage{subfigure}
\usepackage{rotating}

\shorttitle{Strong New Constraints on the EBL}
\shortauthors{Orr et al.}

\begin{document}

\title{Strong New Constraints on the Extragalactic Background Light in the Near- to Mid-IR}

\author{M.R. Orr and F. Krennrich}
\affil{Department of Physics and Astronomy, Iowa State University, Ames, IA 50011}
\email{morr@iastate.edu}

\author{E. Dwek}
\affil{Observational Cosmology Lab, Code 665, NASA Goddard Space Flight Center, Greenbelt, MD 20771}

\begin{abstract}
%% Text of abstract

Direct measurements of the extragalactic background light (EBL) in the near-IR to mid-IR waveband are extremely difficult due to an overwhelming foreground from the zodiacal light that  outshines the faint cosmological diffuse radiation field by more than an order of magnitude.  Indirect constraints on the EBL are provided by  $\gamma$-ray observations of AGN. Using the combination of the \textit{Fermi Gamma-Ray Space Telescope} together with the current generation of ground-based air Cherenkov telescopes  (H.E.S.S., MAGIC, and VERITAS) provides unprecedented sensitivity and spectral coverage for constraining the EBL in the near- to mid-IR.  In this paper we present new limits on the EBL based on the analysis of the broad-band spectra of a select set of $\gamma$-ray blazars covering $200\,$MeV to several TeV.  The EBL intensity at $15\,\mu$m is constrained to be $1.36 \pm 0.58 \, \mathrm{nW} \, \mathrm{m}^{-2} \, \mathrm{sr}^{-1}$.  We find that the fast evolution and baseline EBL models of Stecker et al. (2006), as well as the model of Kneiske et al. (2004), predict significantly higher EBL intensities in the mid-IR ($15\,\mu$m) than is allowed by the constraints derived here.  In addition, the model of Franceschini et al. (2008) and the fiducial model of Dom\'inguez et al. (2011) predict near- to mid-IR ratios smaller than that predicted by our analysis. Namely, their intensities in the near-IR are too low while their intensities in the mid-IR are marginally too high.  All of the aforementioned models are inconsistent with our analysis at the $> \negthickspace 3\sigma$ level. 

\end{abstract}


% \linenumbers

%% main text
\section{Introduction}
\label{Intro}

The extragalactic background light (EBL) contains all radiative energy releases, from nuclear and accretion processes, that have occurred since the epoch of recombination.  While the EBL is the second most dominant cosmological radiation field in our universe, surpassed only by the cosmic microwave background (CMB), it has escaped detection through direct measurements at most wavelengths. Its spectrum is bimodal with one component peaking at $\sim \negthickspace 1 \, \mu\mathrm{m}$,  originating from radiation released through the formation of heavy elements and the accretion of matter onto black holes in active galactic nuclei (AGNs), and a second component peaking at $\sim \negthickspace 100 \, \mu\mathrm{m}$, created through the absorption of UV/optical light that is re-radiated by dust at infrared wavelengths. The energy spectrum contains, therefore, important information about the cosmic evolution of these energy sources, the obscuring dust, and the relative contributions of starburst galaxies and AGNs to the energy released over cosmic time. In addition, the EBL contains information on the rate of core collapse supernovae which provides an important constraint on the normalization of the diffuse supernova neutrino background \citep{Horiuchi:2009}.  The trough in the EBL spectrum at $\sim \negthickspace 15 \, \mu\mathrm{m}$ is caused by the decrease in the stellar and AGN UV-optical emission towards the mid-IR, and the paucity of hot dust capable of emitting at these wavelengths.  This regime is also the most difficult for direct measurements of the EBL due to an overwhelming foreground from zodiacal light.  For detailed reviews of the origin, measurements, modeling, and cosmological implications of the EBL see \citet{Hauser:2001} and \citet{Kashlinsky:2005}.

Very high energy (VHE) gamma-rays ($\sim \negthickspace 0.1 \negthinspace - \negthinspace 10\,$TeV) interact, via pair production, with photons in the near- to mid-IR regime of the EBL (i.e., $\gamma_\text{\tiny{VHE}} \, \gamma_\text{\tiny{EBL}} \rightarrow \mathrm{e}^+ \, \mathrm{e}^-$) \citep{Gould:1967}.  Due to the energy dependence of the  $\gamma$-$\gamma$ optical depth, the observed VHE emission spectra of blazars are softer than the intrinsic spectra. Information regarding the EBL spectral energy distribution (SED) can be gleaned from this spectral absorption using a variety of approaches.  A common technique employed assumes a theoretically based limit on the hardness of the intrinsic spectrum, and hence places a limit on the maximum level of EBL absorption given the softness of the observed spectrum \citep{Aharonian:2006,Aharonian:2007}.  This highest allowed level of absorption translates into an upper limit on the EBL intensity.

Another technique, being an expanded approach to that of \citet{Aharonian:1999}, utilized the TeV spectra of nearby blazars Markarian 421 and Markarian 501 ($z \approx 0.03$) \citep{Dwek:2005}.  The authors identified EBL scenarios producing an exponential rise in luminosity with energy in absorption-corrected spectra.  Such an exponential rise is highly unlikely, based on standard synchrotron self-Compton \citep{Maraschi:1992,Bloom:1996} and external inverse-Compton \citep{Dermer:1993,Sikora:1994} models of blazars, and hence the authors conclude these scenarios are unrealistic.

In this paper we use two different approaches to constrain the EBL, one of which only recently became feasible as a result of \textit{Fermi} measurements of blazar spectra in the 0.1 to 100 GeV energy regime. At these energies, blazar spectra are typically well characterized by a power-law and are a good representations of the intrinsic spectra, since the EBL causes a negligible amount of attenuation at these energies. The first approach, referred to hereafter as the \textit{spectral shape method}, or \textit{Method 1},  assumes that the same power-law characterizing the blazar spectrum at GeV energies extends up to TeV energies, while also allowing for curvature that softens the spectrum at very high energies. Consequently, only EBL SEDs yielding intrinsic blazar spectra consistent with the TeV-extended Fermi power-law are considered viable.  The use of the power-law assumption is, of course, only valid over a limited energy regime since the inverse-Compton (IC) component shows a turnover close to its peak energy.  However, spectra with IC peaks in the multi-TeV regime are generally well characterized by a power-law below the peak \citep{Samuelson:1998,Aharonian:1999,Krennrich:2001,Aharonian:2002}.  Similar approaches have been used by \citet{Georganopoulos:2010} and \citet{Mankuzhiyil:2010} to constrain the EBL absorption optical depth for VHE photons and by \citet{Prandini:2010} to constrain the distances of blazars with unknown redshifts.

The second approach, referred to as the \textit{TeV spectral break method}, or \textit{Method 2}, makes a less stringent assumption, namely that the intrinsic blazar spectrum at TeV energies is characterized by a single power-law, with no restrictions placed on the spectral index. We then search for evidence of a break at $\sim \negthickspace 1\,$TeV in observed blazar spectra, caused by EBL absorption, and compare this with the spectral breaks predicted by different EBL scenarios.  The EBL SEDs producing blazar spectral breaks consistent with observations are considered viable.

We characterize the family of EBL realizations by the $1.6 \, \mu$m and $15 \, \mu$m intensities, and the ratio between the two. As we show later on, the two approaches are complementary. The spectral shape method is most sensitive to the overall EBL intensity, whereas the TeV spectral break method is mostly sensitive to the relative difference between the $1.6 \, \mu$m and $15 \, \mu$m intensities, that is, the slope of the EBL between these two wavelengths. The two methods therefore probe different features of the EBL SED, ruling out different regions of parameter space, providing new constraints on the EBL.  In Section \ref{sec:TestedModels},  we discuss the range of EBL parameter space tested.  Section \ref{sec:AnalysisMethods} describes the two analysis methods used and how they can be combined to improve constraints on the EBL.  The results of our analysis are presented in Section \ref{sec:Results}, with associated interesting and important caveats outlined in Section \ref{sec:Caveats}. Section \ref{sec:ImprovingConstraints} provides some suggestions for improving constraints on the EBL with future blazar measurements.  Finally, a discussion of our results, along with concluding remarks, is provided in Section \ref{sec:DiscussionConclusions}.

\section{EBL  Models}
\label{sec:TestedModels}
A variety of models exist describing the SED of the EBL (e.g., \citet{Kneiske:2002,Primack:2005,Stecker:2006,Franceschini:2008}).  The techniques used to generate these models fall into three basic categories: forward evolution, backward evolution, and observed evolution over a particular redshift range.  The forward evolution method (e.g., \citet{Primack:2005}) makes use of early structure formation scenarios to predict the evolution of galactic luminosity functions forward in time.  Backward evolution techniques (e.g., \citet{Stecker:2006}) begin with the existing galaxy population and model their luminosity functions backward in time.  The third approach to modeling uses deep galaxy surveys (e.g., \citet{Kneiske:2002,Franceschini:2008}) or tracers of chemical evolution to infer the cosmic star formation rate and compute the EBL.

The models tested here follow a more observational approach and were derived from what will be referred to throughout the text as the \textit{baseline model}.  This baseline model follows the shape outlined by lower limits derived from galaxy counts obtained with the \textit{Hubble Space Telescope} \citep{Gardner:2000,Madau:2000}, the \textit{Spitzer Space Telescope} \citep{Fazio:2004,Papovich:2004}, and the \textit{Infrared Space Observatory} \citep{Elbaz:2002}.  An approach similar to that of \citet{Mazin:2007}, using third order splines, was implemented to produce this baseline EBL model.  The advantage to this technique is that it allows one to easily adjust the shape of the SED through manipulation of the control points defining the spline curve.  

\begin{figure}[t]	
	\centering	
	\subfigure{		
		\label{subfig:EBLScenarios}
		\includegraphics[width=3.1in]{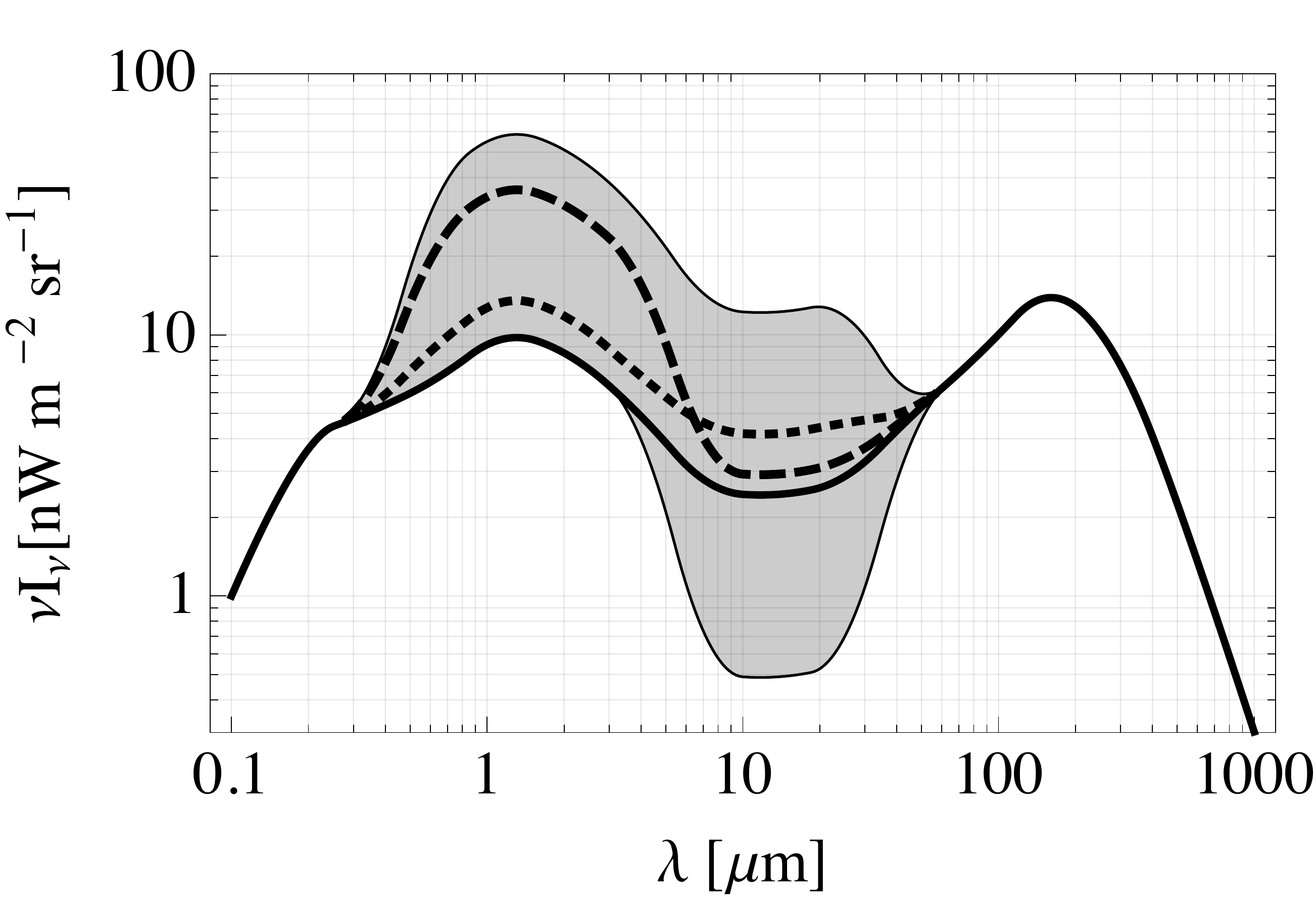}
	}
	\subfigure{		
		\label{subfig:OpticalDepth}
		\includegraphics[width=3.in]{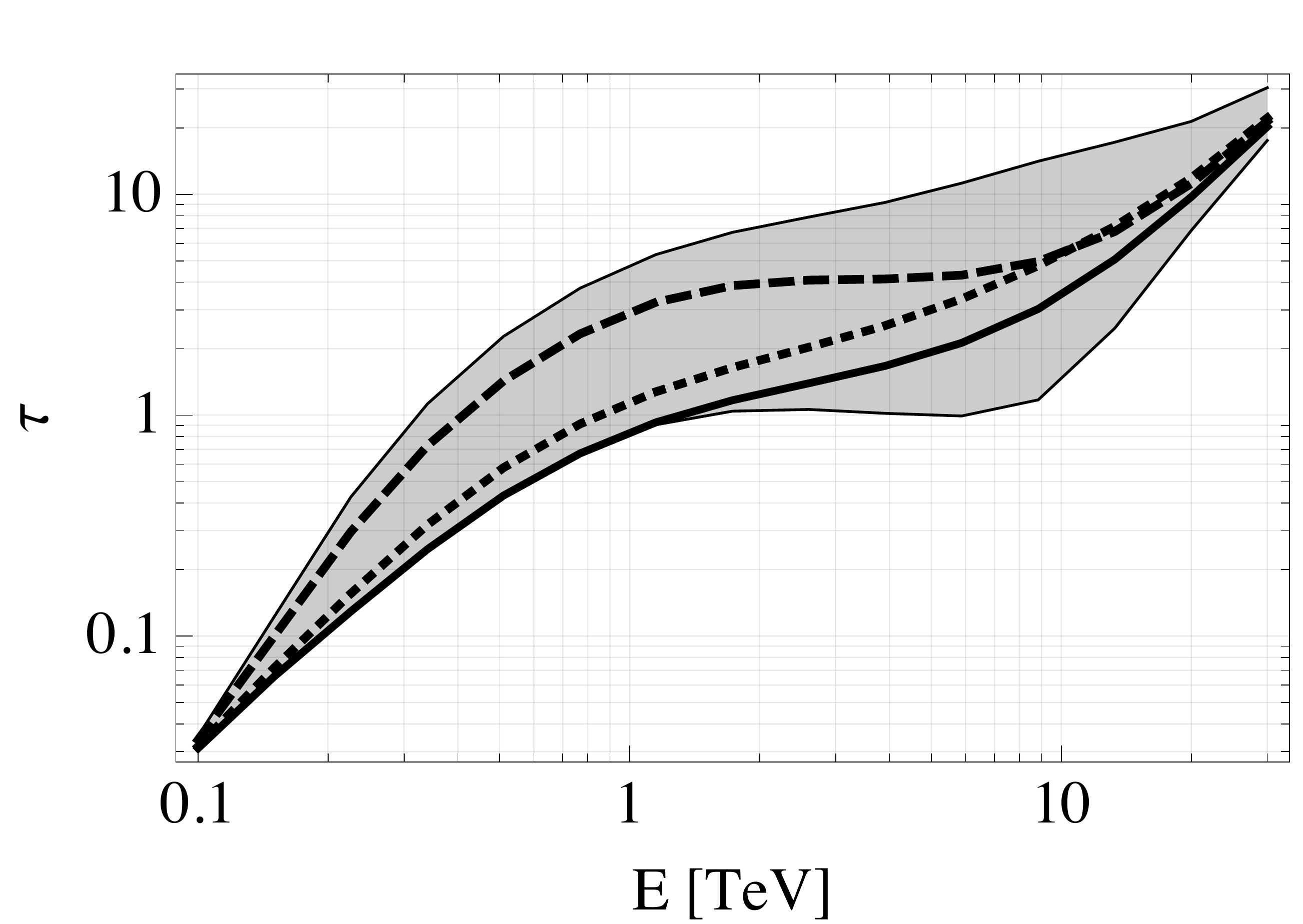}
	}
	\caption{\textit{Left}: EBL intensity versus photon wavelength. The shaded region indicates the range of scenarios tested.  The thick solid line indicates the baseline shape used, from which all other scaled shapes are generated.  For clarity, two additional models are shown (dotted and dashed) illustrating the independent scaling of the near- and mid-IR regions.  \textit{Right}: Optical depth $\tau$ (at $z=0.1$) versus gamma-ray energy in TeV for each EBL scenario tested.  The optical depths for the baseline and two additional EBL models shown in the left panel are shown as well.}
	\label{fig:EBLandTauScenarios}
\end{figure}

Since the absorption of TeV gamma-rays by EBL photons is sensitive to the intensity at near- and mid-IR wavelengths, the regions between $\sim \negthickspace 0.3 \negthinspace - \negthinspace 7 \, \mu\mathrm{m}$ and $\sim \negthickspace 7 \negthinspace - \negthinspace 50 \, \mu\mathrm{m}$ were independently scaled to explore a variety of EBL shapes.\footnote{The near and mid-IR regions were scaled as a whole but future work will investigate scenarios using a finer resolution (e.g, including a near-IR excess due to Population III stars \citep{Dwek:2005b}).} The left panel of Figure \ref{fig:EBLandTauScenarios} illustrates the range of scenarios investigated (shaded region).  The baseline model, indicated by the thick solid curve, is shown along with two additional models (dashed and dotted curves) illustrating various levels of scaling in the two wavelength regimes.  The right panel of Figure \ref{fig:EBLandTauScenarios} shows the calculated optical depths $\tau$ for gamma-rays given each EBL scenario.  All optical depths were calculated assuming the cosmological parameters $H_0 = 70 \, \mathrm{km} \, \mathrm{s}^{-1} \, \mathrm{Mpc}^{-1}$, $\Omega_m = 0.3$, and $\Omega_\Lambda = 0.7$.

To cover the full range of intensities, 18 scaling factors were used in the near-IR regime and 29 in the mid-IR.  This resulted in a total of 519 EBL scenarios.\footnote{The total number of scenarios would have been 522 if not for an imposed restriction that the mid-IR intensity be less than the near-IR intensity.  This resulted in the exclusion of three models.}  The scaling factors were chosen to be linearly distributed in $log_{10}(\nu I_{\nu})$.  

\section{Determining the EBL}
\label{sec:AnalysisMethods}
The analysis methods used here to constrain the EBL are described in Sections \ref{subsec:GeVTeVSpec} and \ref{subsec:SpecBreakTeV}.  In Section \ref{subsec:ContourDiscussion} we describe how these two methods were combined to derive even stronger constraints on the allowable EBL intensity parameter space. 

\subsection{Method 1 - Spectral Shape Method}
\label{subsec:GeVTeVSpec}
With the advent of the \textit{Fermi Large Area Telescope} (LAT) \citep{Atwood:2009}, high energy observations of blazars are now possible in a regime where EBL attenuation is minimal.  The operational energy range of the LAT spans  $\sim \negthickspace 0.1 \negthinspace - \negthinspace 300\,$GeV, overlapping observations with Imaging Atmospheric Cherenkov Telescopes (IACTs) ranging from $\sim \negthickspace 0.05 \negthinspace - \negthinspace 10\,$TeV \citep{Hinton:2004,Albert:2008,Holder:2008}.  Since gamma-rays below $\sim \negthickspace 10\,$GeV travel unimpeded by the EBL \citep{Stecker:2006,Franceschini:2008,Kneiske:2004,Gilmore:2009}, the first method described here used spectra measured by \textit{Fermi} as a proxy for the intrinsic spectra in the TeV regime. Measurements by IACTs were used to calculate a de-absorbed spectrum for each EBL model discussed in Section \ref{sec:TestedModels}.

The intrinsic TeV source spectra were determined using the relationship
\begin{equation}
	\label{eqn:IntrToObsSpec}
	\left( \frac{\mathrm{d}N}{\mathrm{d}E} \right)_\mathrm{intr} = \left( \frac{\mathrm{d}N}{\mathrm{d}E} \right)_\mathrm{obs} e^{\tau(E,z)} ,
\end{equation}
where $(\mathrm{d}N/\mathrm{d}E)_\mathrm{intr}$ is the intrinsic spectrum, $(\mathrm{d}N/\mathrm{d}E)_\mathrm{obs}$ is the observed spectrum, and $\tau(E,z)$ is the optical depth at energy $E$ and source redshift $z$.  The intrinsic spectrum was assumed to have a power-law form given by,\footnote{A power-law is generally a good approximation to blazar spectra over a limited energy range (e.g., $0.1 \, \mathrm{GeV} \leq E \leq 100 \, \mathrm{GeV}$ or $0.1 \, \mathrm{TeV} \leq E \leq 10 \, \mathrm{TeV}$) for measurements with limited statistics.}  
\begin{equation}	
   %\qquad\qquad\qquad\qquad\qquad\quad\;\;~ 
   \left( \frac{\mathrm{d}N}{\mathrm{d}E} \right)_\mathrm{intr} = N_0 \left( \frac{E}{E_0} \right)^{-\Gamma},	
\end{equation}  
where $N_0$ is the normalization at energy $E_0$, $E$ is the energy, and $\Gamma$ is the spectral index. The condition used to evaluate the consistency between the intrinsic TeV spectrum and the extrapolated \textit{Fermi} spectrum was given by
\begin{equation}
\label{eqn:specShapeCriteria}
	\left| \Gamma_\mathrm{TeV} - \Gamma_\mathrm{GeV} \right| \leq N \sqrt{\sigma^2_\mathrm{TeV} + \sigma^2_\mathrm{GeV}},
\end{equation}
where $\Gamma_\mathrm{TeV}$ and $\sigma^2_\mathrm{TeV}$ are the calculated IACT \textit{intrinsic} spectral index and variance, respectively, $\Gamma_\mathrm{GeV}$ and $\sigma^2_\mathrm{GeV}$ are the \textit{Fermi} spectral index and variance, respectively, and $N$ is the confidence level in units of standard deviations.  An EBL scenario was considered viable if the de-absorbed spectrum satisfied the set criteria for consistency with the intrinsic spectrum predicted by the \textit{Fermi} extrapolation.  This criterion was that $\Gamma_\mathrm{TeV}$ be within $\pm N\sigma$ of $\Gamma_\mathrm{GeV}$ (i.e., Equation \ref{eqn:specShapeCriteria}), where $N$ is 1,2, or 3.

As indicated by Equation \ref{eqn:specShapeCriteria}, it was assumed that hard spectrum blazars have no intrinsic spectral break between the \textit{Fermi} and IACT energy regimes up to several TeV.  This assumption was motivated by both observational and theoretical evidence.  Firstly, observations of nearby hard spectrum blazars, such as Mrk 421 and Mrk 501, indicate that their \textit{observed} IC peaks can be located at energies $\gtrsim \negthickspace 1\,$TeV \citep{Dwek:2005,Aleksic:2010}.  Since EBL absorption softens the observed spectrum with respect the the intrinsic one, accounting for any absorption whatsoever will only move this peak to higher energies.  In fact, \citet{Dwek:2005} show that most EBL realizations result in an intrinsic peak energy between $\sim \negthickspace 1$ and 5$\,$TeV for the blazar H 1426+428 ($z=0.129$).  It is also well known that the spectra of Mrk 421 \citep{Krennrich:2001,Aharonian:2002} and Mrk 501 \citep{Samuelson:1998,Aharonian:1999} have exponential cutoffs at energies $\gtrsim \negthickspace 4\,$TeV, further supporting the assumption of an IC peak located at TeV energies.  

Population studies of blazars performed with \textit{Fermi} show that the hardness of the \textit{Fermi} spectrum is correlated with the synchrotron and IC peak frequencies \citep{Abdo:2010:LBAS_SEDs,Abdo:2010:LBAS_SpecProp}.  In other words, a harder Fermi spectrum generally means a higher IC peak energy.  The sources used in Method 1 have some of the hardest spectra of all \textit{Fermi}-detected blazars (even harder than Mrk 421 and Mrk 501) and hence are likely to have some of the highest energy IC peaks.  This provides an additional motivation for the assumption that, for these particular sources, the intrinsic spectrum up to a few TeV is consistent with an extrapolation of the spectrum measured by \textit{Fermi}.  

\citet{Costamante:2003}, \cite{Katarzynski:2006}, \citet{Aharonian:2007:1ES1101}, and \cite{Krennrich:2008} have all shown that extremely hard intrinsic blazar spectra are unavoidable, for these already hard sources, using current EBL model estimates and observationally derived limits.  \citet{Katarzynski:2006} demonstrated that an IC peak at very high energies can be incorporated into the standard synchrotron self-Compton (SSC) emission model of blazars by introducing a low energy cutoff in the parent electron distribution responsible for the synchrotron and SSC emission.  Furthermore, \citet{Tavecchio:2009} showed that recent measurements from \textit{Swift} provide supporting evidence for a very high energy IC peak in the hard spectrum blazar 1ES 0229+200.

\begin{figure}[p]
	\centering	
	\subfigure[1ES 1218+304 (VERITAS \citep{Acciari:2010:1ES1218}).]{		
		\label{subfig:1ES1218}
		\includegraphics[width=3.0in]{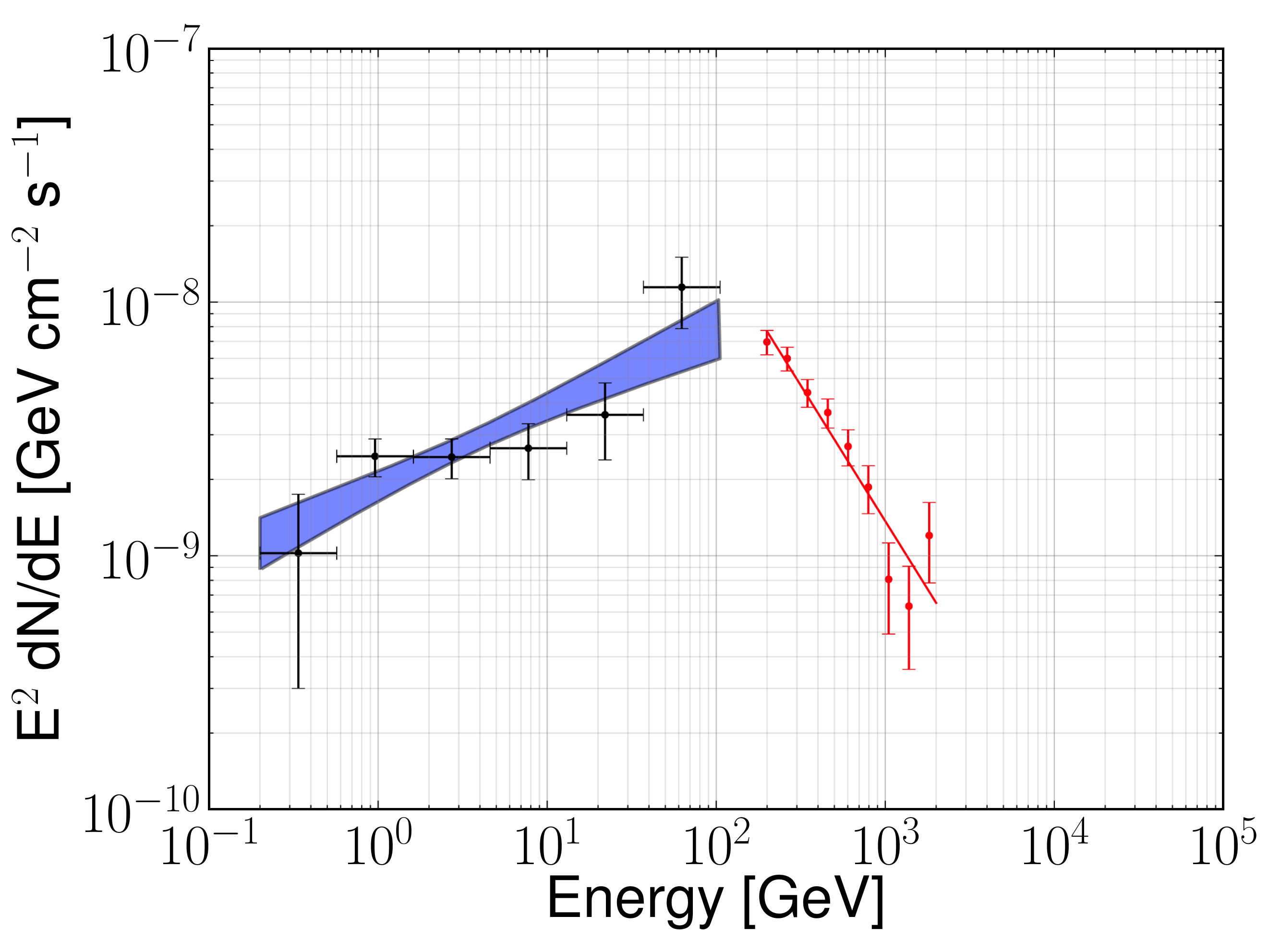}
	}
	\subfigure[1ES 1101-232 (H.E.S.S. \citep{Aharonian:2006}).]{		
		\label{subfig:1E1101}
		\includegraphics[width=3.0in]{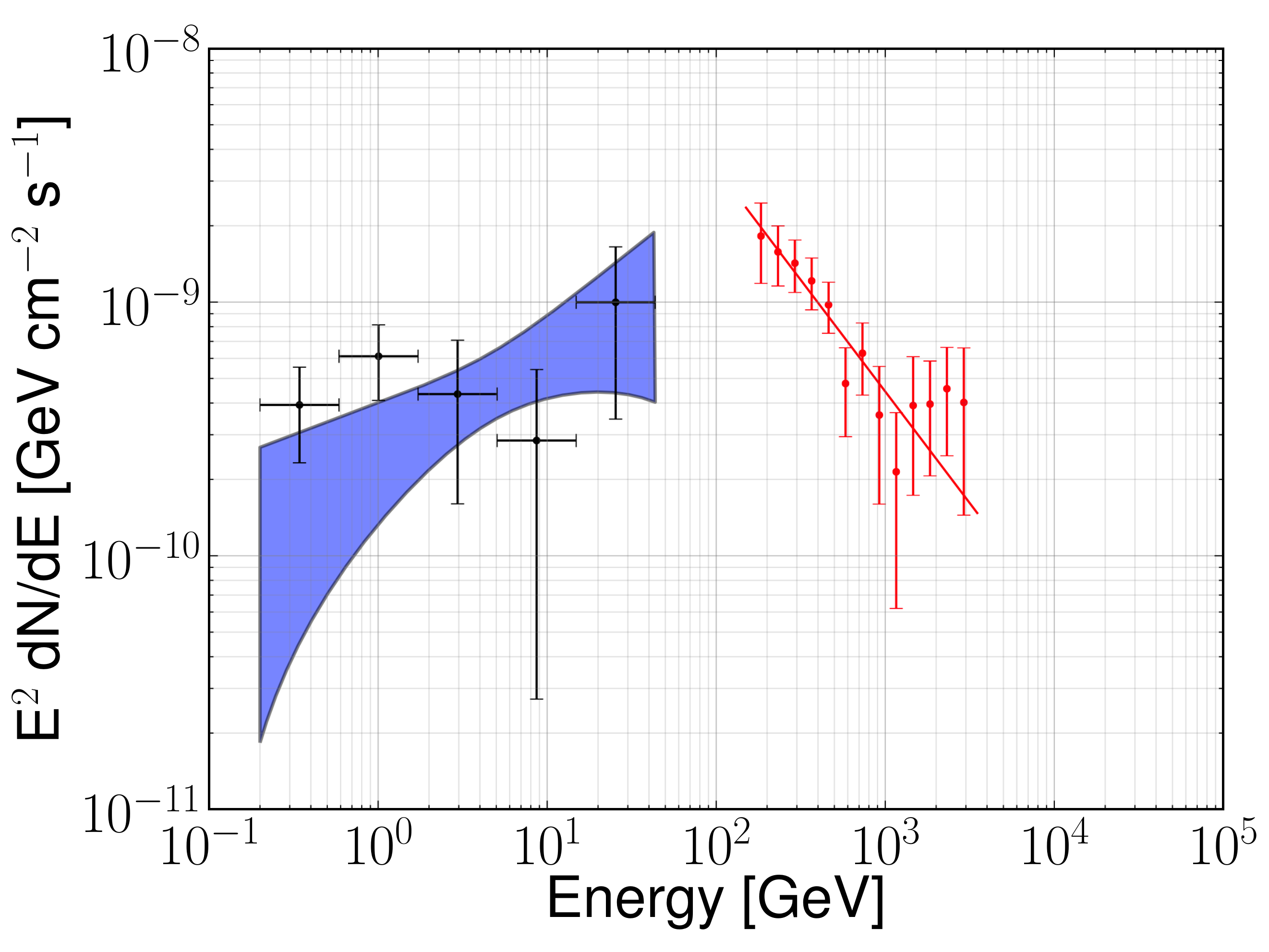}
	}	
	\subfigure[RGB J0710+591 (VERITAS \citep{Acciari:2010:RGBJ0710}).]{		
		\label{subfig:RGBJ0710}
		\includegraphics[width=3.0in]{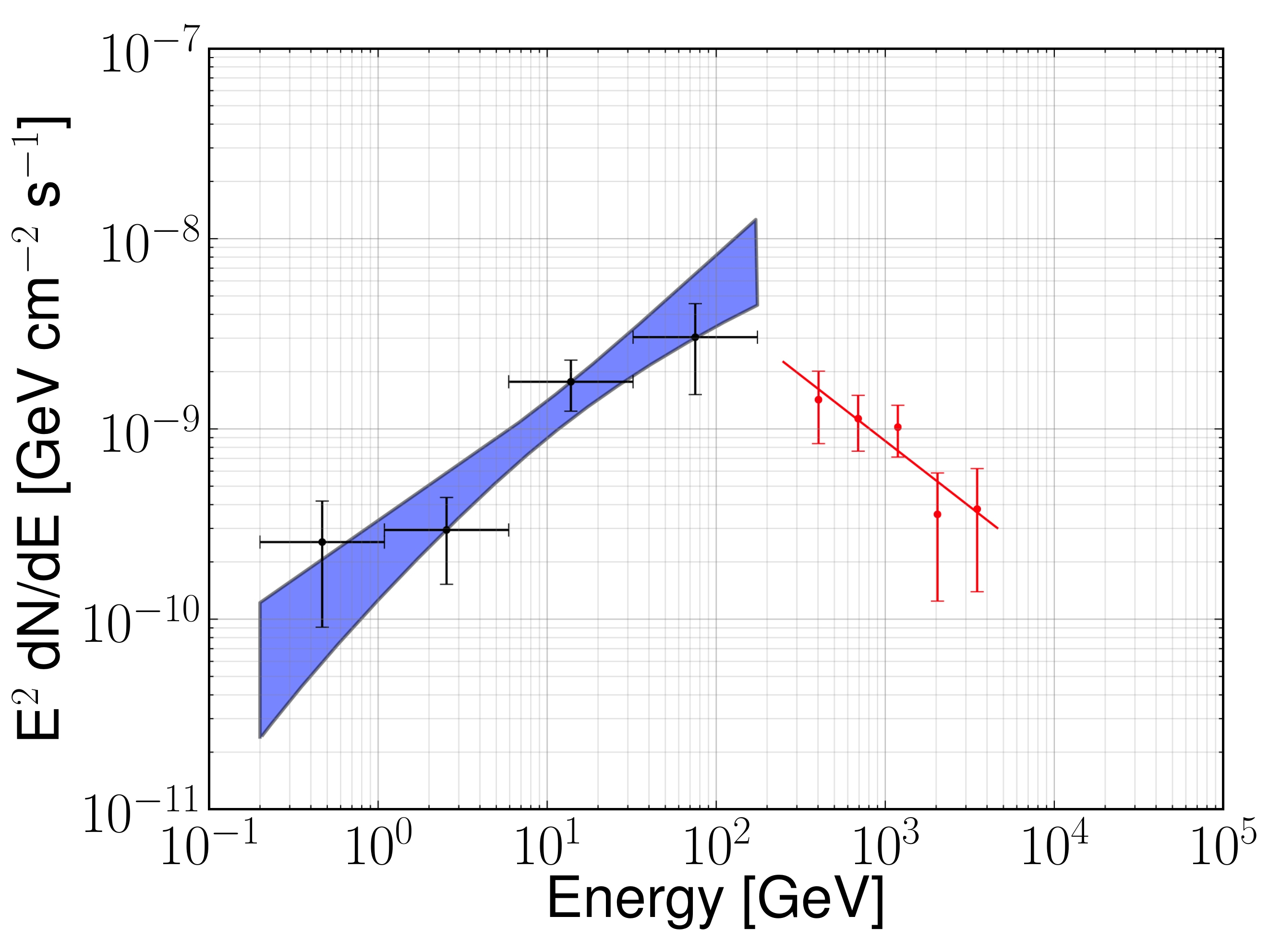}
	}
	\subfigure[1ES 0229+200 (H.E.S.S. \citep{Aharonian:2007}).]{		
		\label{subfig:1ES0229}
		\includegraphics[width=3.0in]{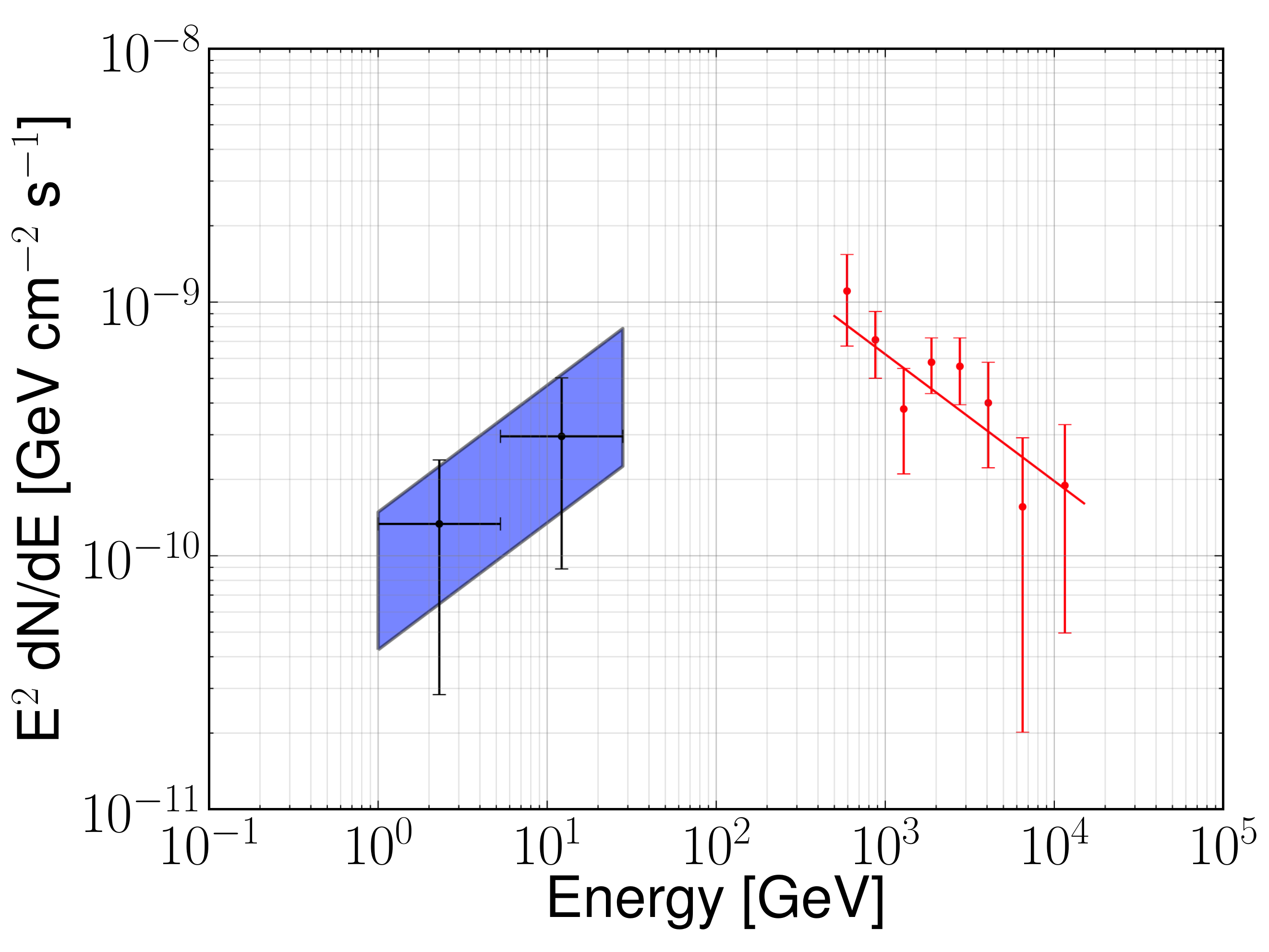}
	}
	\caption{Combined \textit{Fermi} and IACT $E^2 (\mathrm{d}N / \mathrm{d}E)$ spectra in units of $\mathrm{GeV} \, \mathrm{cm}^{-2} \, \mathrm{s}^{-1}$.  The best fit \textit{Fermi} spectrum is indicated by the shaded region.  The \textit{Fermi} flux points were calculated by fixing the spectral index to the value obtained from the fit over the full energy range and then fitting the integral flux over each energy bin.  The IACT spectra are given by the red line and flux points in each plot.  Note: \textit{Fermi} has a weak detection of 1ES 0229+200 (the current result with $\sim \negthickspace 19$ months of data is $\sim \negthickspace 4\sigma$).  For this analysis we have fixed its spectral index in the \textit{Fermi} regime to a value of $1.5$.}
	\label{fig:GeVTeVSpectra}	
\end{figure}

The spectrum of 1ES 0229+200 is the only one used here extending to energies that could potentially span the blazar IC peak for the case of hard spectrum blazars. As such, the analysis of 1ES 0229+200 warranted a slightly different treatment.\footnote{It should be noted that this ``different" treatment was also applied to the other sources used in Method 1.  However, it had no impact on the final results.}  As shown in Figure \ref{subfig:1ES0229}, the spectrum of 1ES 0229+200 \citep{Aharonian:2007} spans the energy range of $\sim \negthickspace 0.5 \negthinspace - \negthinspace 15\,$TeV, nearly 1.5 orders of magnitude.  This broad range, combined with the measurements above several TeV, necessitates the inclusion of a term to account for curvature in the intrinsic spectrum.  This was done using a log-parabolic function of the form
\begin{equation}
	\left( \frac{\mathrm{d}N}{\mathrm{d}E} \right)_\mathrm{intr} = N_0 \left( \frac{E}{E_0} \right)^{-\Gamma -\beta \log_{10}(E/E_0)},   
	\label{eqn:logParabFit}
\end{equation}  
where $N_0$ is the normalization at energy $E_0$, $E$ is the energy, $\Gamma$ is the spectral index, and $\beta$ is the log-parabolic fit parameter.  Additionally, an F-test was performed, along the lines of \citet{Dwek:2005}, to test for the presence of an exponential rise in the intrinsic spectrum at the highest energies.  This exponential fit had the form
\begin{equation}
	\left( \frac{\mathrm{d}N}{\mathrm{d}E} \right)_\mathrm{intr} = \left( \frac{\mathrm{d}N}{\mathrm{d}E} \right)_\mathrm{log} e^{E/\gamma},
\end{equation}
where $( \mathrm{d}N/\mathrm{d}E )_\mathrm{log}$ is the log-parabolic function from Equation \ref{eqn:logParabFit}, $\gamma$ is the exponential fit parameter (with units of energy), and $E$ is the energy.  If the resulting P-value from the F-test was $\geq 95\%$, the exponential function was considered to yield a statistically significant improvement to the fit.  The EBL model was then excluded based on the fact that it produced an unphysical intrinsic spectrum.  A more detailed discussion of the F-test and its application is found in \citet{Dwek:2005}.   

Using the analysis described above, the hard spectrum blazars 1ES 0229+200 \citep{Aharonian:2007}, RGB J0710+591 \citep{Acciari:2010:RGBJ0710}, 1ES 1101-232 \citep{Aharonian:2006}, and 1ES 1218+304 \citep{Acciari:2010:1ES1218} have been studied.    The combined \textit{Fermi} and IACT spectra for these sources are shown in Figure \ref{fig:GeVTeVSpectra}.  It is clear that the spectral breaks between the GeV and TeV regimes, as well as the energy ranges covered, vary between sources.  This results in the derivation of a unique set of EBL constraints for each spectrum (Section \ref{subsec:GeVTeVAnalysis}).  Hard spectrum blazars are ideal for these studies because they are more readily detected at TeV energies \citep{Aharonian:2007,Krennrich:2008}.  They also stand the greatest chance of being detected at energies upwards of 10$\,$TeV, broadening the EBL constraints into the mid-IR regime. Combining the analyses of multiple spectra further constrains the allowable parameter space.  In the following, we refer to this approach as Method 1, or the spectral shape method.

\subsection{Method 2 - TeV Spectral Break Method}
\label{subsec:SpecBreakTeV}
The absorption of gamma-rays by EBL photons may produce breaks in the observed blazar spectrum for some SEDs.  This break is produced by changes in the slope of the $\gamma$-$\gamma$ optical depth.  As illustrated in the right panel of Figure \ref{fig:EBLandTauScenarios}, the optical depth calculated for certain EBL scenarios is nearly flat in the energy range of $\sim \negthickspace 1 \negthinspace - \negthinspace \mathrm{several}$ TeV resulting in an approximately energy independent absorption of gamma-rays.   The observed spectral index in this energy range will be closer to the intrinsic value -- producing a break in the spectrum at $\sim \negthickspace 1\,$TeV.  The magnitude of this break increases with the source redshift and depends on the near- to mid-IR ratio \citep{Imran:2008}.  With currently available data, no individual blazar spectrum is likely to show a statistically significant break in its spectrum.  However, a large sample of blazars may reveal a redshift dependent trend.  The presence/absence of such a trend in observations can be used to constrain the EBL. Table \ref{tab:BlazarSample} lists the sample of blazars used to perform this study.

\begin{sidewaystable}[t]
	\caption{Blazar sample used in the described analyses.  The columns are as follows (left to right): source name, source redshift, GeV and TeV spectral indices, analysis technique performed with the source spectrum, reference to the TeV spectrum, number of TeV spectral points less than (l.t.) and greater than (g.t.) $1.3\,$TeV, and the measured spectral break at $1.3\,$TeV.  All errors given are statistical only.  Method 1 refers to that discussed in Section \ref{subsec:GeVTeVSpec} while 2 refers to Section \ref{subsec:SpecBreakTeV}.  Unless otherwise indicated, $\Gamma_\mathrm{GeV}$ is taken from the \textit{Fermi} 1 Year Catalog (\texttt{http://fermi.gsfc.nasa.gov/ssc/data/access/lat/1yr\_catalog/}).}
	\footnotesize	
	\centering
	\vspace{10pt}	
	\begin{tabular}{c c c c c c c c}
		\hline\hline 
		Source Name & Redshift & \multicolumn{2}{c}{Spectral Index} & Method & Reference & \# Spec. Points  & $\,\,\,\, \Delta\Gamma$ \\
		& & $\Gamma_\mathrm{GeV}$ & $\Gamma_\mathrm{TeV}$ & Used &  & l.t./g.t. $1.3\,$TeV\\
		\hline
		1ES 2344+514 & 0.044 & $1.57 \pm 0.17$ & $2.95 \pm 0.12$ & 2 & \citet{Albert:2007} & 4/3 & $-0.39 \pm 1.11$ \\
		1ES 1959+650 & 0.048 & $2.10 \pm 0.05$ & $2.58 \pm 0.18$ & 2 & \citet{Tagliaferri:2008} & 4/2 & $ -0.61 \pm 1.01$ \\
		PKS 0548-322 & 0.069 & - & $2.8 \pm 0.3$ & 2 & \citet{Aharonian:2010} & 3/2 & $\,\,\,\, 1.75 \pm 0.83$ \\
		PKS 2005-489 & 0.071 & $1.90 \pm 0.06$ & $4.0 \pm 0.4~$ & 2 & \citet{Aharonian:2005:PKS2005} & 6/3 & $\,\,\,\, 1.93 \pm 2.30$ \\
		RGB J0152+017 & 0.080 & - & $2.95 \pm 0.36$ & 2 & \citet{Aharonian:2008} & 4/2 & $\,\,\,\, 0.96 \pm 1.18$ \\ 
		PKS 2155-304$^{\dagger}$ & 0.117 & $1.91 \pm 0.02$ & $3.32 \pm 0.06$ & 2 & \citet{Aharonian:2005:PKS2155} & 7/3 & $\,\,\,\, 0.30 \pm 0.40$ \\
		RGB J0710+591 & 0.125 & $\,\,\, 1.30 \pm 0.16^*$ & $2.69 \pm 0.26$ & 1,2 & \citet{Acciari:2010:RGBJ0710} & 3/2 & $-0.96 \pm 1.14$ \\
		H 1426+428 & 0.129 & $1.49 \pm 0.18$ & $3.50 \pm 0.35$ & 2 & \citet{Petry:2002} & 3/4 & $\,\,\,\, 2.11 \pm 0.55$ \\
		1ES 0229+200 & 0.140 & $\,\,\,\,\, 1.50 \pm 0.20^{*\ddagger}$ & $2.50 \pm 0.19$ & 1,2 & \citet{Aharonian:2007} & 3/5 & $\,\,\,\, 0.39 \pm 0.62$ \\
		1ES 1218+304 & 0.182 & $\,\,\, 1.69 \pm 0.07^*$ & $3.07 \pm 0.09$ & 1,2 & \citet{Acciari:2010:1ES1218} & 7/2 & $\,\,\,\, 1.33 \pm 1.35$ \\
		1ES 1101-232 & 0.186 & $\,\,\, 1.61 \pm 0.26^*$ & $2.88 \pm 0.17$ & 1,2 & \citet{Aharonian:2006} & 9/4 & $\,\,\,\, 1.64 \pm 0.64$ \\
		1ES 0347-121 & 0.188 & - & $3.10 \pm 0.23$ & 2 & \citet{Aharonian:2007:1ES0347} & 4/3 & $\,\,\,\, 1.28 \pm 0.89$ \\
    	\hline\hline
	\end{tabular}
	\begin{tabular}{c c c c c c c c}
		* & \multicolumn{7}{l}{\textit{Fermi} analysis performed in this work.} \\
    	$\dagger$ & \multicolumn{7}{l}{$< \negthickspace 1\%$ probability of being a steady \textit{Fermi} source.} \\
    	$\ddagger$ & \multicolumn{7}{l}{\textit{Fermi} has a weak detection of 1ES 0229+200 at $\sim \negthickspace 4\sigma$.  For this analysis we have \textit{assumed} a \textit{Fermi} spectral index of $1.5 \pm 0.2$.}
	\end{tabular}	
	\label{tab:BlazarSample}
\end{sidewaystable}

The shape of the EBL uniquely determines the spectral break versus redshift distribution and was calculated here for a multitude of scenarios using a test blazar spectrum.  This test spectrum was chosen to represent the ``average" spectrum of the blazars listed in Table \ref{tab:BlazarSample}.  This was motivated, for example, by the fact that the spectrum of 1ES 0229+200 \citep{Aharonian:2007} ranges from $\sim \negthickspace 0.5 \negthinspace - \negthinspace 15\,$TeV whereas that of 1ES 1218+304 \citep{Acciari:2010:1ES1218} spans $\sim \negthickspace 0.2 \negthinspace - \negthinspace 2\,$TeV.  Using a test blazar spectrum ranging from $0.2 \negthinspace - \negthinspace 15\,$TeV would be unrealistic as there is no individual source in the sample covering such a wide energy range.  To generate a spectrum representative of the overall sample of blazars, the average lowest energy bin, highest energy bin, and number of bins in the sample were calculated.  This resulted in a test spectrum characterized by 8 energy bins ranging from 200$\,$GeV to 5$\,$TeV.  The flux normalization and spectral index (intrinsic values) of the test blazar spectrum are irrelevant since only the magnitude of the spectral break is of concern.  Nevertheless, values need to be chosen to perform the analysis.  The Crab Nebula flux at 1 TeV was used for the normalization along with an intrinsic spectral index of 1.5.  The underlying assumption in this method is that the intrinsic blazar spectrum is well described by a single power-law over the energy range considered.  

Beginning with an exact power-law function, and calculating the spectrum due to EBL absorption, one is left with an observed spectrum that shows  detailed features of the $\gamma$-$\gamma$ optical depth and is decidedly not of a power-law (or broken power-law) form. This is, of course, not what is observed by IACTs.  To produce an absorbed spectrum more consistent with observations, each flux point was assigned an error of 25\% of the flux in that bin.  This value was chosen to approximately coincide with the mean error of all flux points from the IACT measurements in the blazar sample.  Gaussian fluctuations were added to the spectral points using a Normal distribution with a standard deviation equal to the 25\% error bars.  Optical depths for the test spectrum were calculated for all EBL scenarios over the redshift range 0.05--0.45 in steps of 0.05.  No EBL evolution with redshift was implemented.  

The absorbed spectra were calculated using the inverse relation of Equation \ref{eqn:IntrToObsSpec}.  Each spectrum was fit with a broken power-law of the form
\begin{equation}
	\frac{\mathrm{d}N}{\mathrm{d}E} = 
	\begin{cases}
		N_0 \left( \dfrac{E}{E_\mathrm{break}} \right)^{-\Gamma_1} & , ~ E \leq E_\mathrm{break} \\ \\
		N_0 \left( \dfrac{E}{E_\mathrm{break}} \right)^{-\Gamma_2} & , ~ E > E_\mathrm{break}
	\end{cases} ~,
	\label{eqn:BrknPwrLaw}
\end{equation}
where $N_0$ is the normalization at the break energy $E_\mathrm{break}$, $\Gamma_1$ and $\Gamma_2$ are the spectral indices below and above $E_\mathrm{break}$, respectively, and $E$ is the energy.  The spectral break was defined as
\begin{equation}
	\Delta\Gamma = \Gamma_1 - \Gamma_2 ~ .
\end{equation} 
With this definition, a spectral hardening above the break energy would result in a positive $\Delta\Gamma$.  From here the distribution of $\Delta\Gamma$ as a function of redshift was determined for each EBL scenario.  This process was repeated 100 times, mimicking multiple independent observations, each time yielding a different result due to the Gaussian fluctuations in the spectra.  This facilitated the determination of the break versus redshift distribution to an arbitrary precision and accuracy.  

Each distribution was fit with a line, i.e., 
\begin{equation}
\label{eqn:BreakVsZFit}
	\Delta\Gamma(z) = m \, z + b ~,
\end{equation}
where $\Delta\Gamma(z)$ is the spectral break at redshift $z$ and $m$ and $b$ are free parameters.  Using the best linear fit obtained from observational data, the $1\sigma$, $2\sigma$, and $3\sigma$ two dimensional contours in slope and intercept space (i.e., $m$ and $b$, respectively, in Equation \ref{eqn:BreakVsZFit}) were determined.  The best fit results from the expected spectral break versus redshift distribution for each EBL scenario were then compared with the observationally derived contours.  If the fit parameters for a particular scenario fell within the two dimensional $1 \sigma$ contour, this scenario was included as part of the $1\sigma$ contour in EBL parameter space.  The $2\sigma$ and $3\sigma$ contours in EBL parameter space were determined in the same fashion (Section \ref{subsec:BreakAnalysis}).  In the following, we refer to this approach as Method 2, or the TeV spectral break method.

\subsection{Complementarity of Methods 1 \& 2 in Constraining the EBL}
\label{subsec:ContourDiscussion}
\begin{figure}[t]	
	\centering	
	\subfigure[]{	
		\includegraphics[width=3.1in]{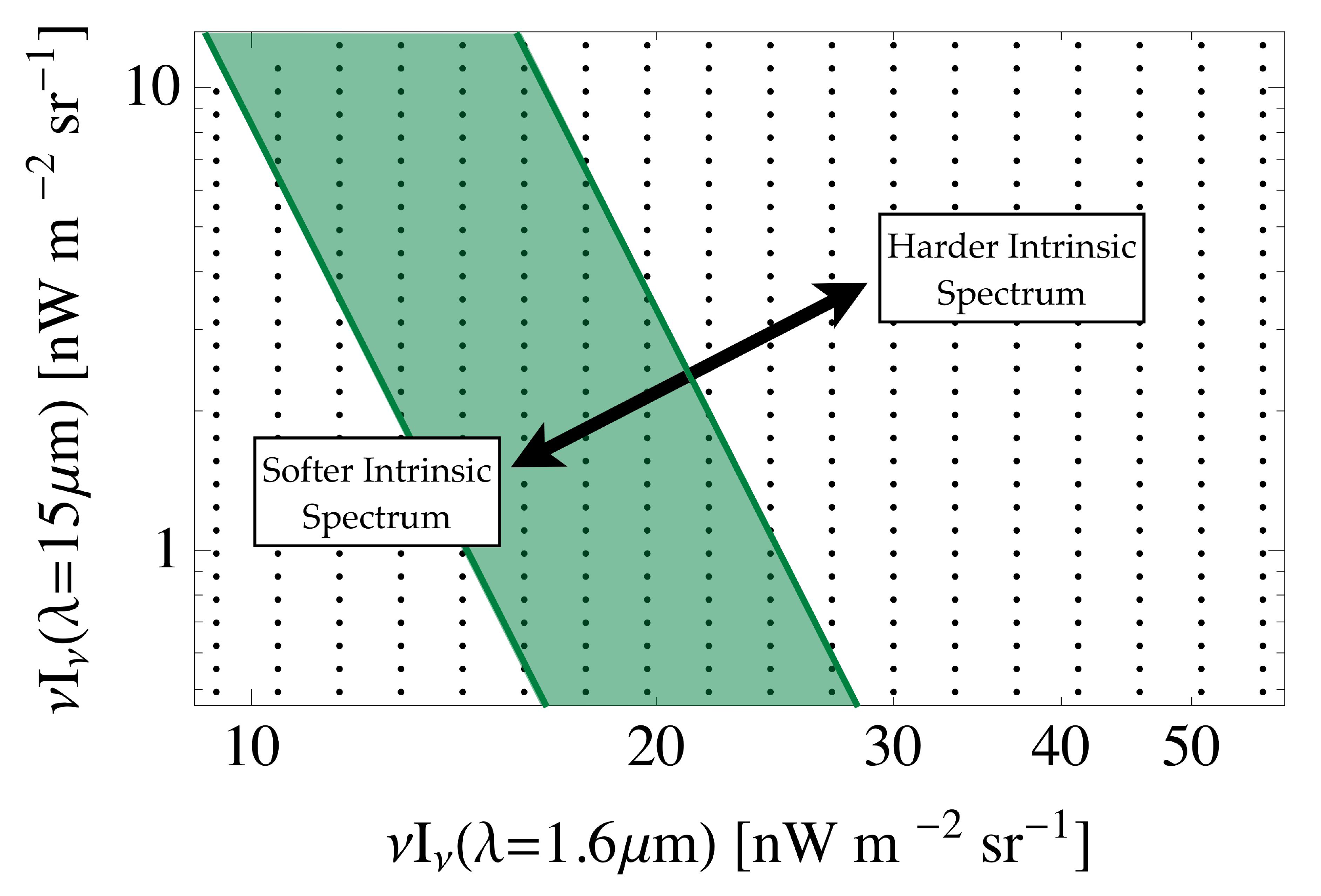}
		\label{subfig:ContourIllustration-GeVTeVBreak}
	}
	\subfigure[]{
		\includegraphics[width=3.1in]{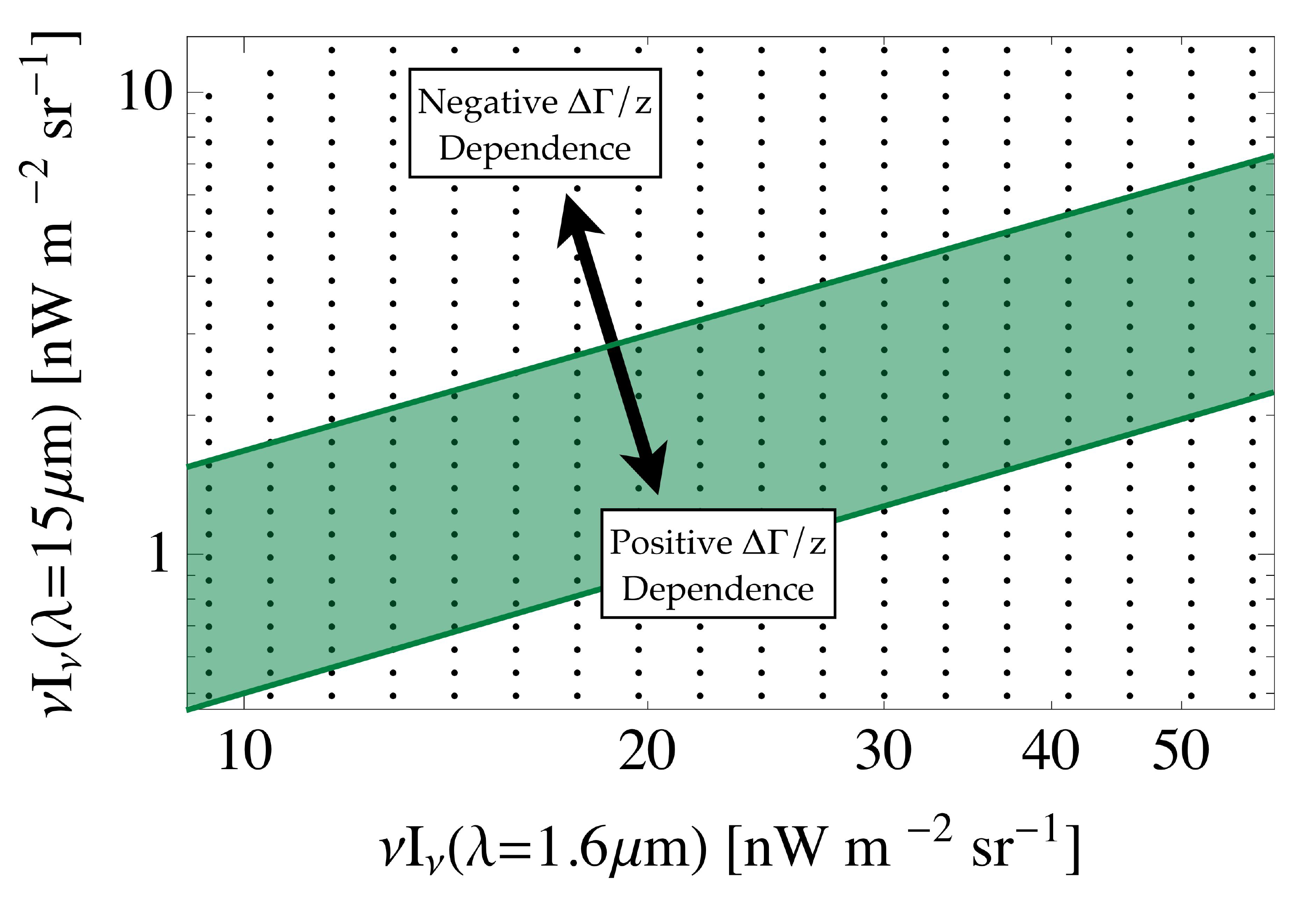}			
		\label{subfig:ContourIllustration-TeVBreak}		
	}	
	\caption{EBL intensity at $15 \, \mu \mathrm{m}$ plotted versus the intensity at $1.6 \, \mu \mathrm{m}$.  Each grey dot represents an EBL scenario tested. (a) As the intensities at $1.6 \, \mu \mathrm{m}$ and $15 \, \mu \mathrm{m}$ increase (moving up and to the right on the plot), the inferred intrinsic blazar spectrum becomes harder.  The intrinsic spectrum softens as the two intensities decrease (down and to the left).  This is indicated by the arrows in the figure. By placing restrictions on the maximum hardness and softness of the intrinsic spectrum, one obtains a contour in intensity space oriented approximately perpendicular to these two directions (shaded region).  (b) Increasing the near-IR to mid-IR ratio increases the redshift dependence of the EBL-induced spectral break at TeV energies (i.e., $\Delta\Gamma(z)/z$ becomes larger).  Decreasing this ratio results in an increasingly negative dependence of spectral break versus redshift (i.e., $\Delta\Gamma(z)/z$ becomes smaller to the point of being negative).  This is indicated by the arrows in the figure.  By comparing the redshift dependence of the TeV spectral break produced by various EBL models with observations, one obtains a contour in intensity space oriented as shown by the shaded region.}
	\label{fig:ContourIllustration}	
\end{figure}

The constraints placed on the EBL are presented in the form of a contour plot.  This contour is drawn in the parameter space defined by the near-IR intensity at $1.6 \, \mu\mathrm{m}$ and the mid-IR intensity at $15 \, \mu\mathrm{m}$.  These two wavelengths are representative of the positions of the near-IR peak and mid-IR trough in the EBL SED.  Given that the EBL models tested are relatively smooth, these two parameters serve as an adequate characterization of a given EBL scenario.  An illustration of such a contour plot is shown in Figure \ref{fig:ContourIllustration}.  Each point represents one of the EBL scenarios tested.  Shown in this fashion, the linear sampling of intensity in log-space is clearly seen.  A constant ratio of intensities (i.e., $\nu I_{\nu}(1.6 \, \mu \mathrm{m}) / \nu I_{\nu}(15 \, \mu \mathrm{m})$) would be given by a straight line with positive slope.  It should be noted that parametrizing the results in this fashion does not mean the derived constraints only apply to the EBL intensity at $1.6 \, \mu\mathrm{m}$ and $15 \, \mu\mathrm{m}$.  This presentation is simply a way of characterizing the average slope of the EBL SED between these two wavelengths.

The shaded region in Figure \ref{subfig:ContourIllustration-GeVTeVBreak} shows how a contour, derived from the analysis outlined in Section \ref{subsec:GeVTeVSpec}, might appear.  As the near- and/or mid-IR intensities increase (decrease), the inferred intrinsic spectrum hardens (softens).  A contour oriented as in Figure \ref{subfig:ContourIllustration-GeVTeVBreak} is obtained when placing limitations on the maximum hardness/softness of the intrinsic spectrum.

An approximately orthogonal contour is obtained from the analysis outlined in Section \ref{subsec:SpecBreakTeV}.  This is because, when increasing the near-IR to mid-IR intensity ratio, the dependence of the EBL induced spectral break with redshift becomes increasingly positive.  That is to say, $\Delta\Gamma(z)/z$ becomes larger.  Conversely, decreasing the near-IR to mid-IR ratio makes $\Delta\Gamma(z)/z$ smaller to the point where it can become negative.  These directions are shown on the contour illustration in Figure \ref{subfig:ContourIllustration-TeVBreak}.  By comparing the redshift dependence of the TeV spectral break produced by different EBL models with that from observations, one can obtain a contour in EBL intensity space as represented in Figure \ref{subfig:ContourIllustration-TeVBreak} by the shaded region.

Comparing the two contours from Figure \ref{fig:ContourIllustration}, it becomes clear that the EBL constraining methods outlined in Sections \ref{subsec:GeVTeVSpec} and \ref{subsec:SpecBreakTeV} are complementary to one another.  Each method is sensitive to different characteristics of the EBL SED thereby improving the constraints at near- and mid-IR wavelengths.

\section{Results}
\label{sec:Results}
The techniques described in Sections \ref{subsec:GeVTeVSpec} and \ref{subsec:SpecBreakTeV} have been applied to a sample of blazars (Table \ref{tab:BlazarSample}).  The four blazars selected for the spectral shape analysis were chosen for their redshifts ($z > 0.1$), hard spectra, and steady emission in the \textit{Fermi} regime.    

The blazar sample used in the TeV spectral break analysis was selected based on characteristics of the source spectra in the TeV regime.  Most importantly, to test for a spectral break at $\sim \negthickspace 1\,$TeV, each spectrum must have at least two data points both above and below the break energy.  Spectra from large flaring events were excluded as their atypically high statistics skew the overall results.  An example of this was the $\sim \negthickspace 7\,$Crab flare event of PKS 2155-304 \citep{Aharonian:2007:PKS2155Flare}.  The two nearby blazars Mrk 421 and Mrk 501 were also excluded from this analysis due to the fact that the presence of a cutoff at $\sim \negthickspace 4\,$TeV in their intrinsic spectra \citep{Krennrich:2001,Samuelson:1998}, combined with their low redshift ($z \approx 0.03$), greatly inhibits the search for a spectral break resulting from EBL absorption.

The analysis results for Methods 1 and 2 are presented in Sections \ref{subsec:GeVTeVAnalysis} and \ref{subsec:BreakAnalysis}, respectively, and are then combined in Section \ref{subsec:CombinedAnalyses}.  

\subsection{Spectral Shape Analysis}
\label{subsec:GeVTeVAnalysis}
The source spectra used for the spectral shape analysis are shown in Figure \ref{fig:GeVTeVSpectra}.  Each \textit{Fermi} spectrum consists of approximately 19 months of data and was analyzed using the \texttt{Science Tools v9r15p2} package distributed by the \textit{Fermi} collaboration.\footnote{http://fermi.gsfc.nasa.gov/ssc/data/analysis/software/}  Only events with zenith angles less than $105^{\circ}$ were included to avoid contamination from Earth's gamma-ray albedo.  All exposures and fluxes were calculated using the post-launch instrument response function \texttt{P6\_V3\_DIFFUSE}.  The galactic and extragalactic diffuse gamma-ray backgrounds were modeled using data available on the \textit{Fermi} Science Support Center website.\footnote{http://fermi.gsfc.nasa.gov/ssc/data/access/lat/BackgroundModels.html}  The variability index of each source, as quoted in the \textit{Fermi} 1 Year Catalog\footnote{http://fermi.gsfc.nasa.gov/ssc/data/access/lat/1yr\_catalog/}, is less than 23.21 indicating a $> \negthickspace 99\%$ probability the source emission is steady.  Spectral fitting was performed from $200\,$MeV to the highest photon energy associated with the given source.  TeV spectra were obtained from the references listed in Table \ref{tab:BlazarSample}.

\begin{figure}[p]	
	\centering	
	\subfigure[1ES 1218+304]{		
		\label{subfig:1ES1218-Contours}
		\includegraphics[width=3.in]{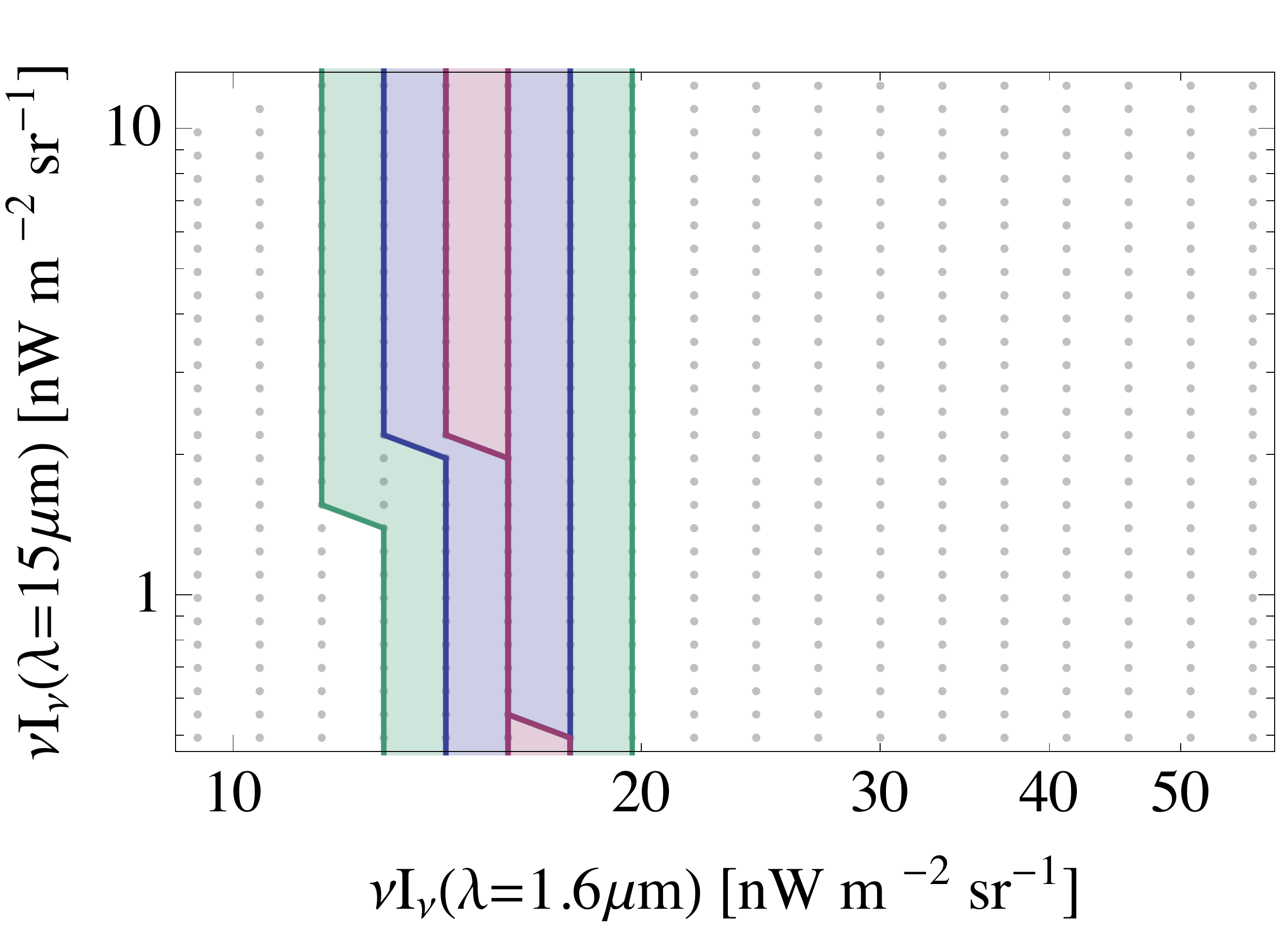}
	}	
	\subfigure[1ES 1101-232]{		
		\label{subfig:1ES1101-Contours}
		\includegraphics[width=3.in]{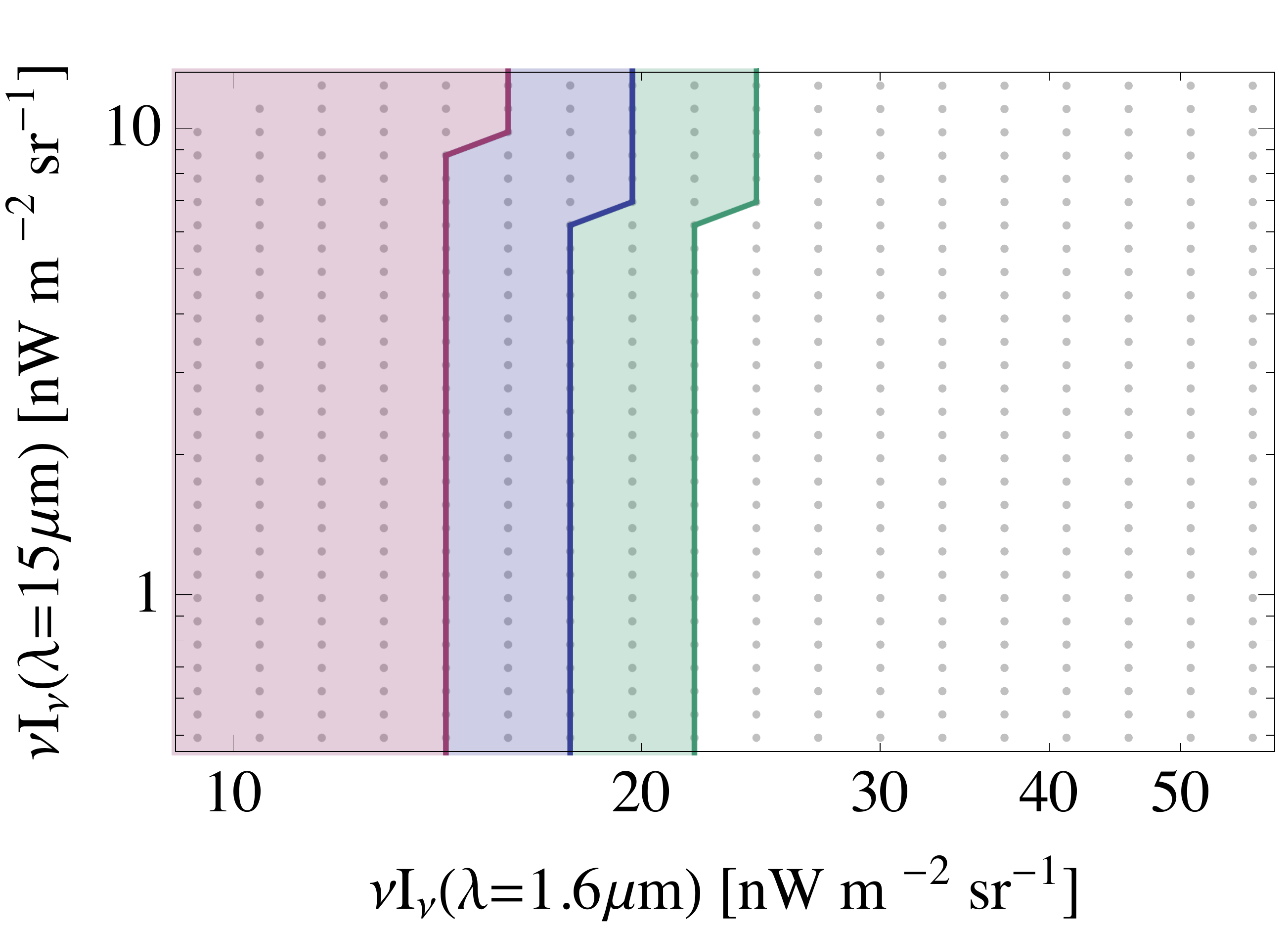}
	}	
	\subfigure[RGB J0710+591]{		
		\label{subfig:RGBJ0710-Contours}
		\includegraphics[width=3.in]{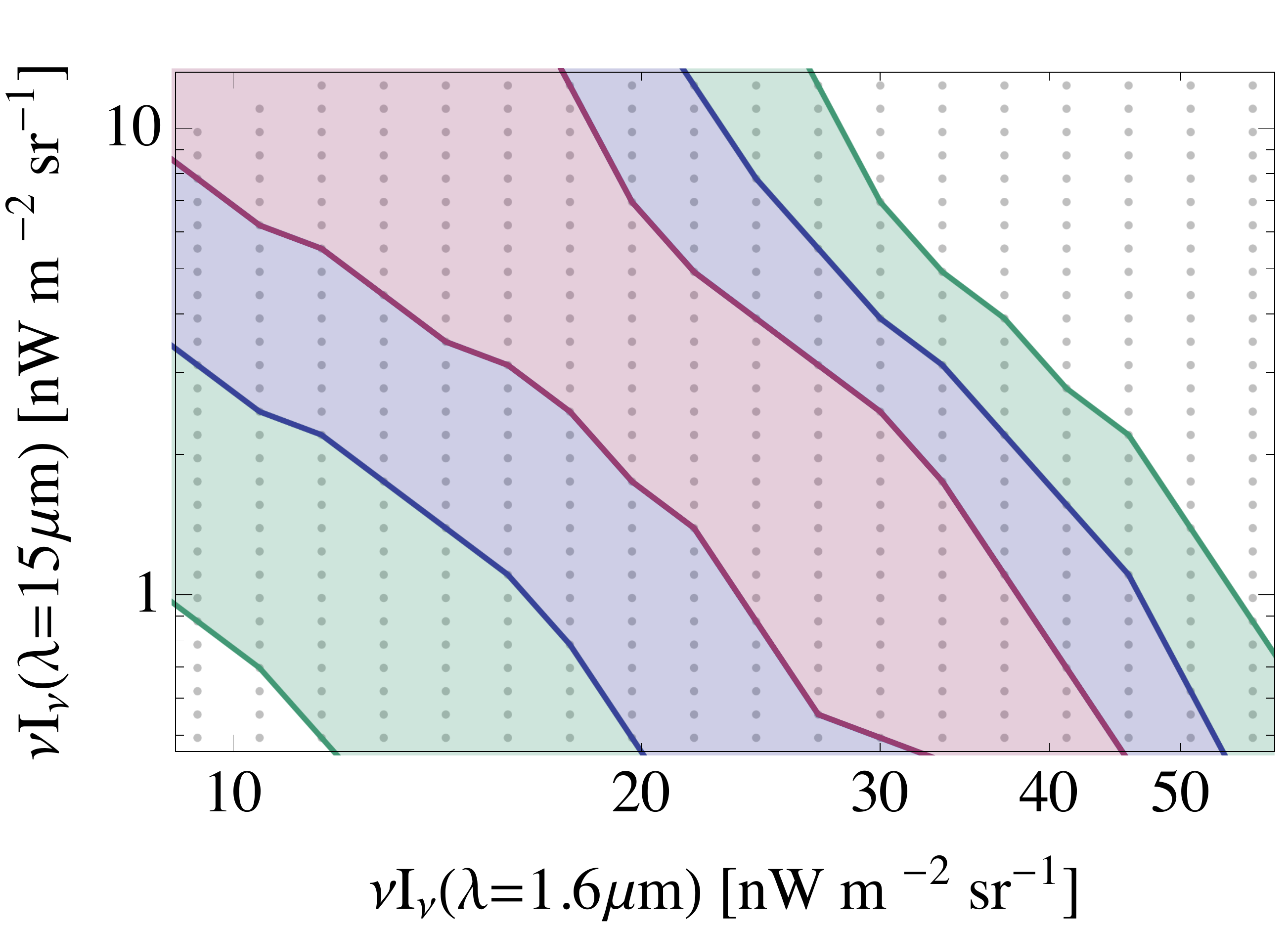}
	}
	\subfigure[1ES 0229+200]{		
		\label{subfig:1ES0229-Contours}
		\includegraphics[width=3.in]{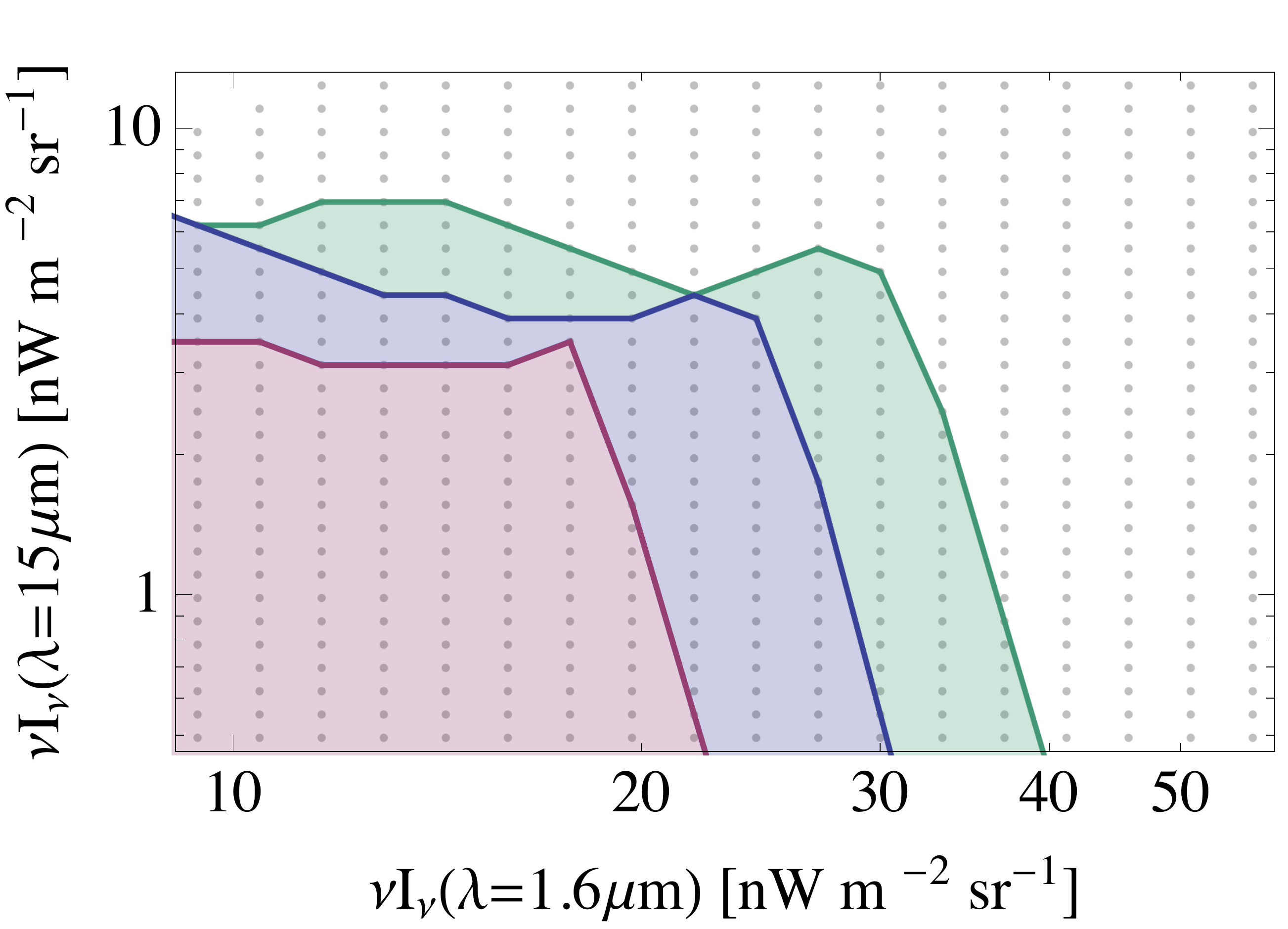}
	}
	\caption{Constraints, provided by Method 1, on the EBL intensity, in units of $\mathrm{nW} \, \mathrm{m}^{-2} \, \mathrm{sr}^{-1}$, at $1.6\,\mu$m and $15\,\mu$m.  The contours shown are for the 1 (red), 2 (blue), and 3 (green) sigma confidence intervals.  Each grey dot represents one of the EBL scenarios tested.  Note: \textit{Fermi} has not obtained a significant detection of 1ES 0229+200.  For this analysis we have assumed the spectral index in the \textit{Fermi} regime to be $1.5 \pm 0.2$.}
	\label{fig:EBL-BlazarContours}
\end{figure}

The \textit{Fermi} analysis of 1ES 0229+200 was handled differently than for the other three sources due to the fact that \textit{Fermi} has only a weak detection of this source.  Nevertheless, it was included since it has the hardest TeV spectrum of all the blazars studied here and provides a unique set of constraints.  To obtain a spectrum from the \textit{Fermi} data, an assumed spectral index of $1.5 \pm 0.2$ was used.  This parameter was fixed in the source model so that the the best fit flux normalization could be determined.  The overall test statistic (TS) for this source, between 1 and 28$\,$GeV, was 18.0 ($\sim \negthickspace 4 \sigma$).  The TS values for the first ($1 \negthickspace - \negthickspace 5\,$GeV) and second ($5 \negthickspace - \negthickspace 28\,$GeV) energy bins were 2.6 and 15.4, respectively.

The $1\sigma$, $2\sigma$, and $3\sigma$ contours for each blazar, constraining the EBL intensity at $1.6\,\mu$m and $15\,\mu$m are shown in Figure \ref{fig:EBL-BlazarContours}.  The reader is cautioned not to view these contours as strict exclusion regions for other authors' models since the analysis is dependent on the shape of the EBL SED.  However, given that most models have the same general shape, it is a safe assumption that scenarios falling outside the $3\sigma$ contours derived here are not consistent with this analysis.  

Figure \ref{fig:EBL-BlazarContours} shows that gamma-ray observations provide strong constraints on the $1.6 \, \mu$m EBL intensity, but leave the $15 \, \mu$m intensity largely undetermined.  This is due to the lack of TeV data extending up to multi-TeV energies which provide the strongest constraints on the mid-IR region of the EBL.  It is worthwhile to point out some particular properties of the different contours in Figure \ref{fig:EBL-BlazarContours} as they provide insight into the constraints derived with each blazar. 

One of the most noticeable features is the narrowness of the 1ES 1218+304 contours (Figure \ref{fig:EBL-BlazarContours}(a)).  This is a result of the high precision measurement of the spectral index by \textit{Fermi} and VERITAS in both the GeV and TeV regimes (see Table \ref{tab:BlazarSample}).   The measurement errors are a factor of 2 or more smaller than those of the other blazars in the sample.  Another distinguishing feature is the vertical orientation of the contours indicating that the TeV spectrum is minimally affected by EBL absorption in the mid-IR regime.  This results from the fact that the spectrum only extends up to $\sim \negthickspace 2\,$TeV.  

The spectrum of 1ES 1101-232 extends up to $\sim \negthickspace 3\,$TeV but is also insensitive to the mid-IR portion of the EBL (Figure \ref{fig:EBL-BlazarContours}(b)).  This is most likely due to the somewhat large error on the spectral index measured by \textit{Fermi} ($1.87 \pm 0.28$) as well as the large power-law fit probability ($\sim \negthickspace 80\%$) to the observed TeV spectrum (see \citet{Aharonian:2006}).\footnote{A fit probability of 80\% suggests the possibility of either overestimated errors or somewhat correlated data points.  In either case, fluctuations in the highest energy spectral points (those sensitive to the mid-IR portion of the EBL) will have a limited impact on the overall power-law fit to the de-absorbed spectrum.}

In contrast to 1ES 1218+304 and 1ES 1101-232, the contours for RGB J0710+591 (Figure \ref{fig:EBL-BlazarContours}(c)) are somewhat inclined, indicating sensitivity to the mid-IR EBL intensity.  The spectrum of RGB J0710+591 extends up to $\sim \negthickspace 4\,$TeV demonstrating that a modest increase in the maximum energy of the spectrum can have a significant impact on EBL studies.  

The spectrum of 1ES 0229+200 is unique in that it starts at $\sim \negthickspace 500\,$GeV and extends up to $\sim \negthickspace 15\,$TeV.  As a result, this source is the most sensitive, of all those analyzed here, to absorption by the mid-IR portion of the EBL.  This is evident in Figure \ref{fig:EBL-BlazarContours}(d).  While none of the aforementioned blazars strongly constrain the $15 \, \mu$m EBL intensity, the spectral shape analysis of 1ES 0229+200 constrains the intensity to be $\lesssim \negthickspace 5 \, \mathrm{nW} \, \mathrm{m}^{-2} \, \mathrm{sr}^{-1}$.

\subsection{TeV Spectral Break Analysis}
\label{subsec:BreakAnalysis}
The energy dependence of the $\gamma$-$\gamma$ optical depth for some EBL scenarios may produce a break around $\sim \negthickspace 1\,$TeV in blazar spectra (Section \ref{subsec:SpecBreakTeV}).  Here, a spectral break energy of $1.3\,$TeV is assumed.  The choice of this value is easily arrived at using the the fact that the $\gamma \negthickspace - \negthickspace \gamma$ cross section peaks at $\lambda_\epsilon (\mu \mathrm{m}) \approx 1.24 E_\gamma (\mathrm{TeV})$, where $\lambda_\epsilon$ is the EBL photon wavelength and $E_\gamma$ is the gamma-ray energy \citep{Dwek:2005}.  Since the near-IR peak of the EBL occurs at $\lambda_\epsilon \approx 1.6 \, \mu$m, a flattening of the $\gamma$-$\gamma$ optical depth would be expected at $E_\gamma \approx 1.3\,$TeV.  The right panel of Figure \ref{fig:EBLandTauScenarios} shows that the optical depth does, in fact, flatten out between $1 \negthinspace - \negthinspace 2\,$TeV, further motivating this choice for the spectral break energy.

\begin{figure}[t]		
	\centering	
	\hspace{0.125in}	
	\includegraphics[width=6.in]{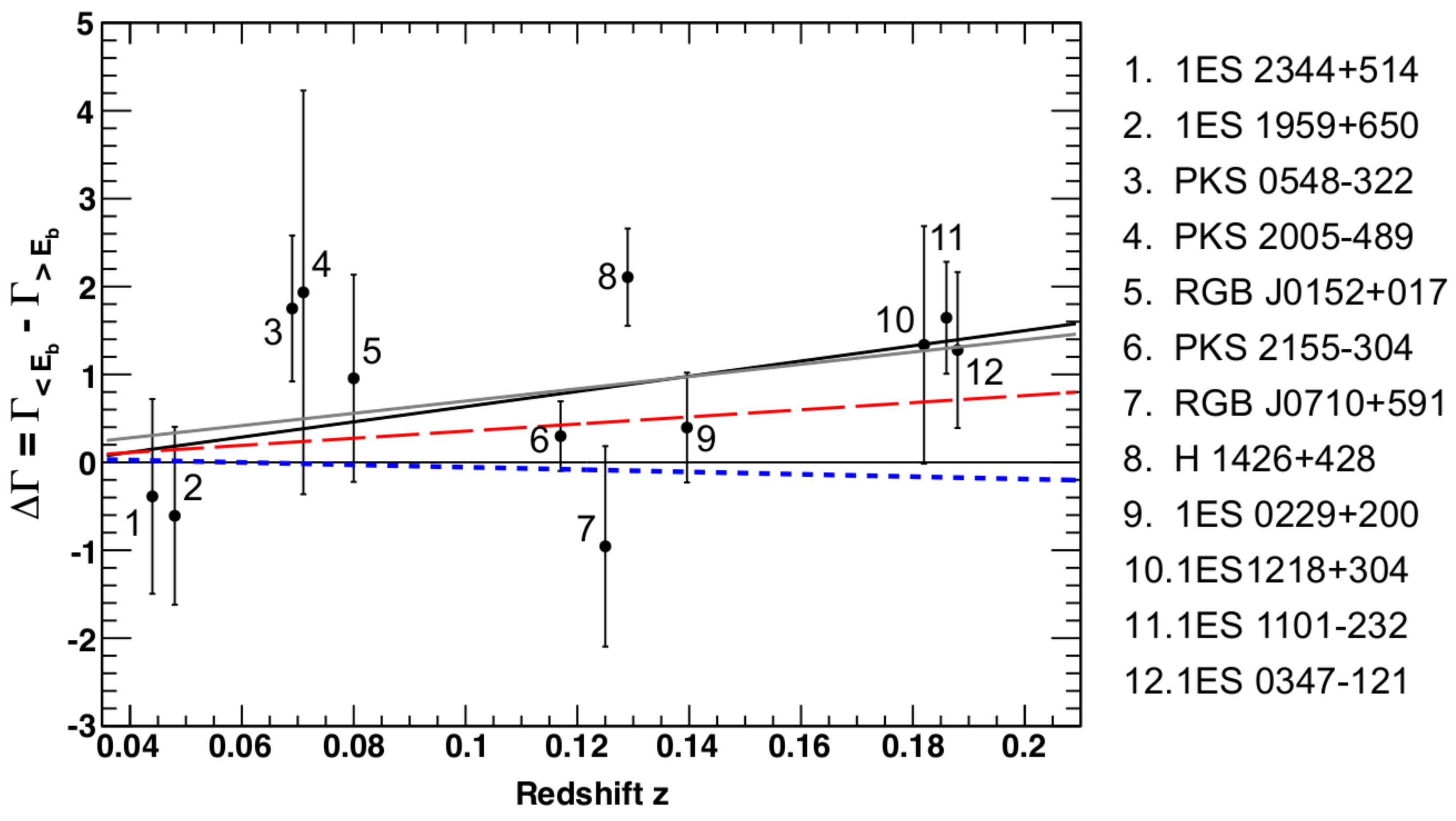}
	\caption{Distribution of the observed spectral break versus redshift assuming a break energy of 1.3$\,$TeV (the choice of which is motivated by the near-IR peak at $\sim \negthickspace 1 \, \mu$m in the EBL scenarios used here).  The sample includes the 12 blazars (see Table \ref{tab:BlazarSample}) available to date with spectra extending above and below the break energy. The spectral break is defined as the spectral index below the break energy $E_\mathrm{b}$ minus the spectral index above the break energy (i.e., $\Delta\Gamma = \Gamma_{<\mathrm{E_b}} - \Gamma_{>\mathrm{E_b}}$) The black solid line represents the best linear fit given by $\Delta\Gamma(z) = (8.68 \pm 5.37)z - (0.24 \pm 0.71)$ and yields $\chi^2/\nu = 14.10/10$.  The grey solid line shows the best linear fit assuming $\Delta\Gamma(0) = 0$, giving $\Delta\Gamma(z) = (6.95 \pm 1.65)z$ with $\chi^2/\nu = 14.22/11$.  The long-dashed line shows the $\Delta\Gamma(z)$ distribution expected from the scenario of \citet{Aharonian:2006}, scaled by a factor of 0.55, yielding $\chi^2/\nu = 18.09/12$.  The short-dashed line shows the $\Delta\Gamma(z)$ distribution expected from the model of \citet{Franceschini:2008}, yielding $\chi^2/\nu = 36.23/12$.  Fitting the data  with a flat line at $\Delta\Gamma(z) = 0$ yields $\chi^2/\nu = 31.98/12$.}
	\label{fig:TeVBreakVsZ}
\end{figure}

To determine the spectral break versus redshift distribution ($\Delta\Gamma(z)$), a sample of 12 blazars (Table \ref{tab:BlazarSample}) with spectra extending both above and below the break energy were fit with a broken power-law (Equation \ref{eqn:BrknPwrLaw}).  Figure \ref{fig:TeVBreakVsZ} shows the distribution of $\Delta\Gamma(z)$ versus the source redshift for all blazars in the sample. Two linear fits were performed on the data -- one leaving the intercept $\Delta\Gamma(0)$ as a free parameter and one with the intercept fixed to 0.  The results of these two fits were
\begin{equation}
	\Delta\Gamma(z) = 
	\begin{cases}
		(8.68 \pm 5.37)z - (0.24 \pm 0.71) & , ~ \chi^2/\nu = 14.10/10 \\
		(6.95 \pm 1.65)z & , ~ \chi^2/\nu = 14.22/11
	\end{cases} .
\end{equation}
In the absence of any spectral break one would expect $\Delta\Gamma(z) = 0$.  The data exclude this hypothesis at the $\sim \negthickspace 3.2 \sigma$ level.  

Figure \ref{fig:TeVBreakVsZ} also shows the $\Delta\Gamma(z)$ distributions produced by the models of \citet{Aharonian:2006},\footnote{The EBL SEDs from \citet{Aharonian:2006} are scaled versions of the model from \citet{Primack:2001}.} scaled by a factor of 0.55, and \citet{Franceschini:2008}.  A summary of how well these models describe the data, along with the other authors' models considered here, is given in Table \ref{tab:TeVBreakAnalysisSummary}.  Column 3 gives the best linear fit to the $\Delta\Gamma(z)$ distribution produced by the given model. This is derived in the same way outlined in Section \ref{subsec:SpecBreakTeV}.  However, in this case we use the EBL model from the reference listed rather than one of our own realizations discussed in Section \ref{sec:TestedModels}.  With the exception of the EBL scenario from \citep{Aharonian:2006}, all models are inconsistent with TeV spectral break data at the $> \negthickspace 3\sigma$ level.

\begin{table}[t]
	\caption{Summary of the comparison between TeV spectral break data and the EBL models from other authors considered here.  The columns are as follows: \textit{Column 1}: EBL model name. \textit{Column 2}: Reference. \textit{Column3}: Best linear fit, of the form $\Delta\Gamma(z) = mz +b$, to the TeV spectral break distribution produced by the model (i.e., simulated, not observed data). \textit{Column 4}: $\chi^2$ of the observed data assuming the best fit from Column 3. \textit{Column 5}: Confidence level with which the model can be excluded given the data. Note: Parenthetical numbers in Columns 4 and 5 are obtained using the technique discussed in Section \ref{subsec:SpecBreakCaveats}.}
	\footnotesize	
	\centering
	\vspace{10pt}	
	\begin{tabular}{c c c c c}
		\hline\hline 
		Model & Reference & ~~ Best Fit $\Delta\Gamma(z)$ & $~\, \chi^2$ & $1-P(\chi^2,\nu=12)$ \\
		\hline
		Aharonian & \citet{Aharonian:2006} & $~~\, 7.41 z - 0.10$ & 14.16 (22.95) & $1.1\sigma$ ($2.2\sigma$) \\
		Aharonian $\times ~ 0.55$ & \citet{Aharonian:2006} & $~~\, 4.05 z - 0.05$ & 18.09 (21.97) & $1.6\sigma$ ($2.1\sigma$) \\
		Aharonian $\times ~ 0.45$ & \citet{Aharonian:2006} & $~~\, 3.32 z + 0.00$ & 19.06 (23.24) & $1.7\sigma$ ($2.2\sigma$) \\
		Stecker (fast) & \citet{Stecker:2006} & $-2.83 z + 0.42$ & 31.66 (48.54) & $3.2\sigma$ ($4.7\sigma$) \\
		$\Delta\Gamma(z) = 0$ & N/A & $~~\, 0.00 z + 0.00$ & 31.98 & $3.2\sigma$ \\
		Dominguez & \citet{Dominguez:2010} & $-0.94 z + 0.07$ & 34.46 (31.49) & $3.4\sigma$ ($3.2\sigma$) \\
		Franceschini & \citet{Franceschini:2008} & $-1.35 z + 0.08$ & 36.23 (32.67) & $3.6\sigma$ ($3.3\sigma$) \\		
		Kneiske & \citet{Kneiske:2004} & $-1.18 z - 0.05$ & 36.46 (34.54) & $3.6\sigma$ ($3.5\sigma$) \\ 
		Stecker (baseline) & \citet{Stecker:2006} & $-4.29 z - 0.02$ & 61.87 (52.86) & $5.7\sigma$ ($5.1\sigma$) \\
		\hline\hline
	\end{tabular}
	\label{tab:TeVBreakAnalysisSummary}
\end{table}

\begin{figure}[t]	
	\centering	
	\includegraphics[width=5.25in]{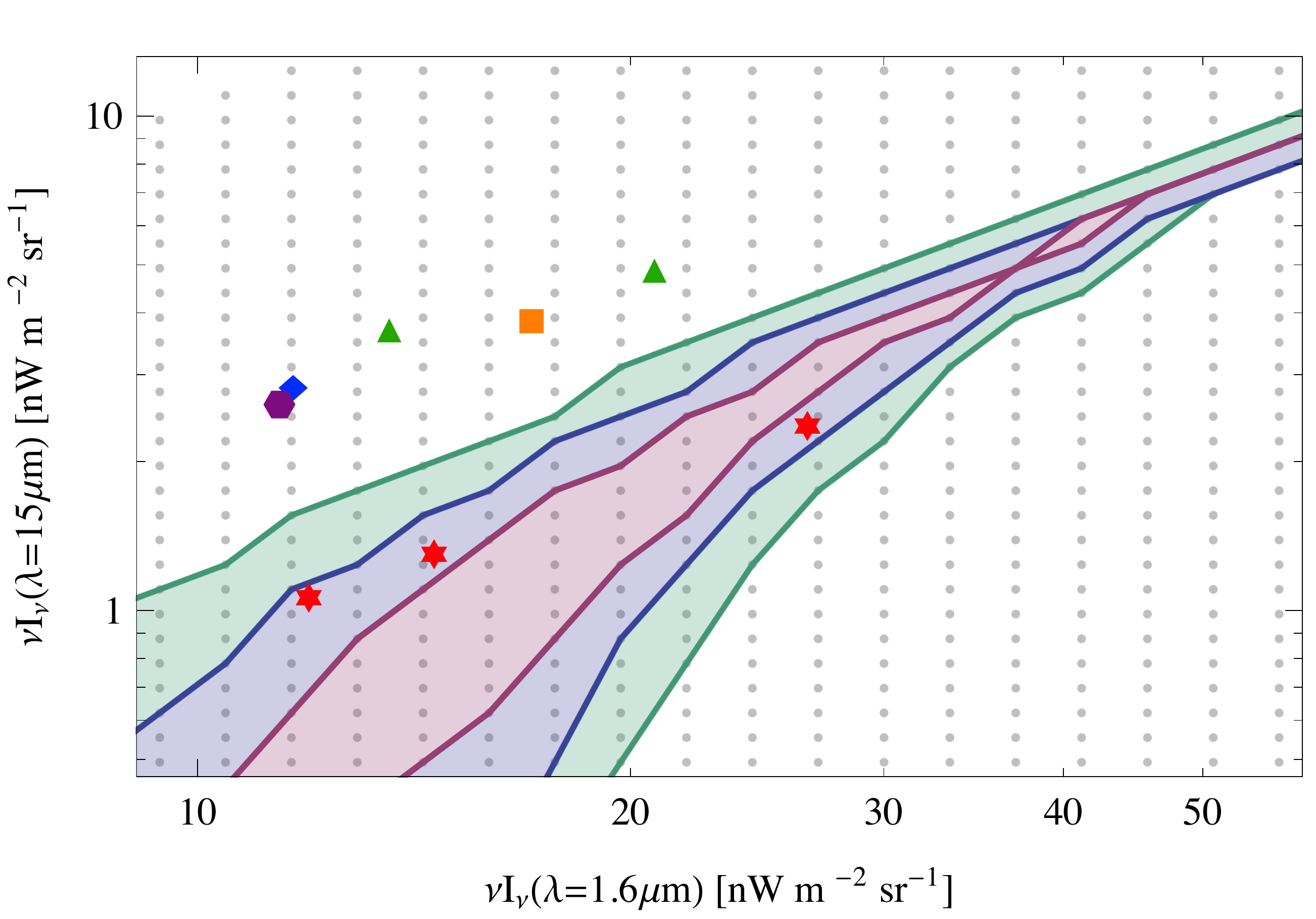}
	\caption{Constraints, provided by Method 2, on the EBL intensity, in units of $\mathrm{nW} \, \mathrm{m}^{-2} \, \mathrm{sr}^{-1}$, at $1.6\,\mu$m and $15\,\mu$m.  The contours shown are for the 1 (red), 2 (blue), and 3 (green) sigma confidence intervals. The intensities for the models of \citet{Aharonian:2006} (red stars), \citet{Stecker:2006} (green triangles), \citet{Franceschini:2008} (blue diamond), \citet{Kneiske:2004} (orange square), and \citet{Dominguez:2010} (purple hexagon) are also shown.  The three red stars for the scenario of \citet{Aharonian:2006} represent scaling factors of 1.0, 0.55, and 0.45 (right to left).  For the model of \citet{Stecker:2006}, the two green triangles represent the so called fast and baseline evolution models (right to left).}
	\label{fig:EBL-TeVBreakContours}
\end{figure}

Figure \ref{fig:EBL-TeVBreakContours} shows the $1\sigma$, $2\sigma$, and $3\sigma$ contours resulting from comparison between the $\Delta\Gamma(z)$ distributions generated from the EBL models tested and observations (Figure \ref{fig:TeVBreakVsZ}).  In addition, points taken from \citet{Aharonian:2006}, \citet{Kneiske:2004}, \citet{Stecker:2006}, \citet{Franceschini:2008}, and \citet{Dominguez:2010} show where these other EBL models fall within the contours generated here.\footnote{It should be noted that some of these EBL models have bumps at $\sim \negthickspace 15 \, \mu$m.  The data points should therefore only be used as a point of reference and not for a strict comparison.  The compatibility of these models with our results was determined using the full EBL SED.}  The orientation of the contours are as anticipated from the discussion in Section \ref{subsec:ContourDiscussion}.  The limits on the near- to mid-IR ratio ($\nu I_{\nu}(1.6 \, \mu \mathrm{m}) / \nu I_{\nu}(15 \, \mu \mathrm{m})$) become increasingly restrictive as the intensities at $1.6\,\mu$m and $15\,\mu$m increase.  Using this technique, future direct measurements of the EBL at near-IR wavelengths (for instance by the \textit{James Webb Space Telescope}\footnote{http://www.jwst.nasa.gov/}) will help to further constrain the EBL in the mid-IR regime.

\subsection{Combining the Analyses}
\label{subsec:CombinedAnalyses}
\begin{figure}[t]	
	\centering	
	\subfigure[Overlayed $2\sigma$ contours.]{		
		\label{subfig:EBL-Overlayed2SigmaContours}
		\includegraphics[width=3.1in]{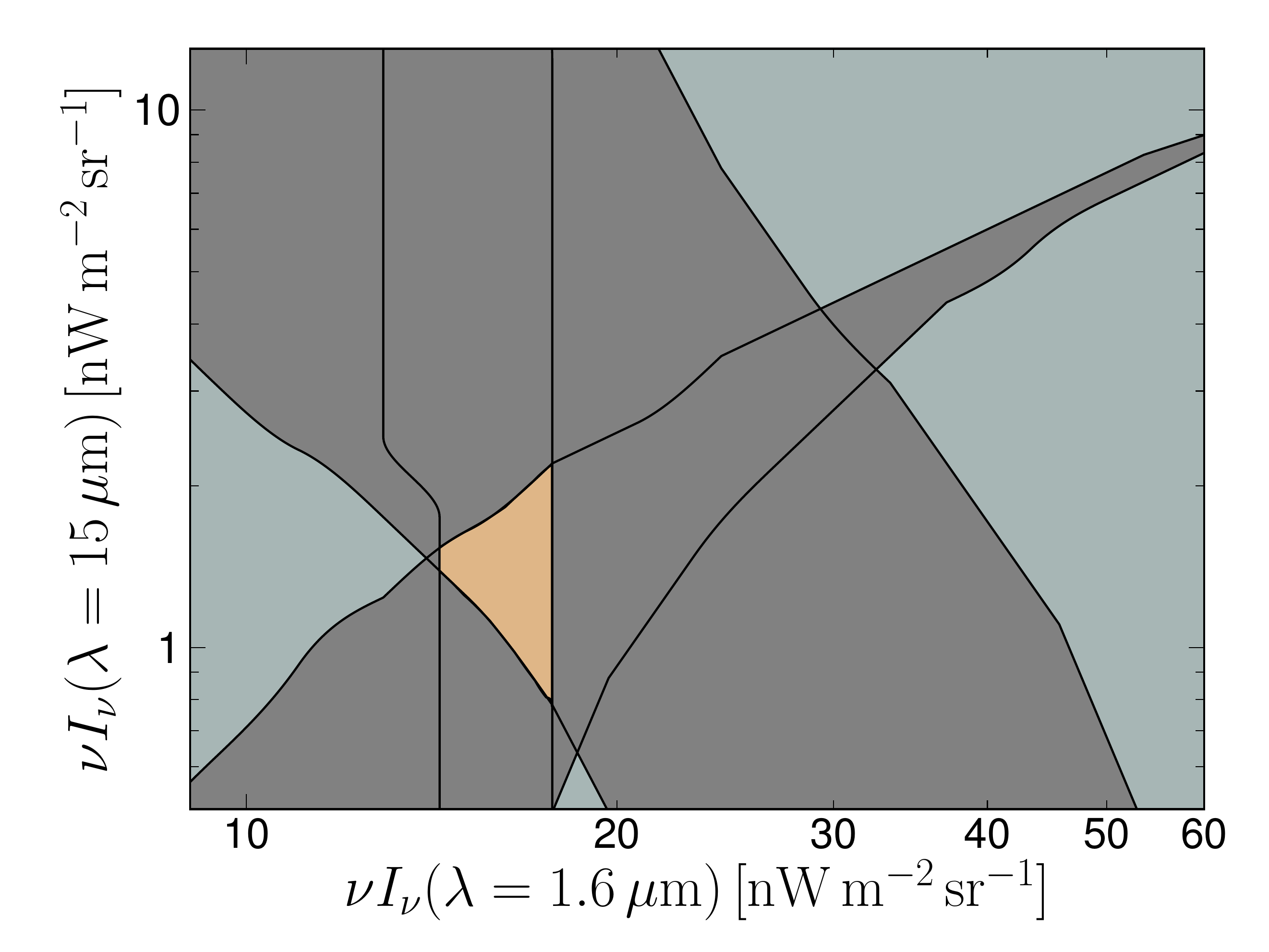}
	}
	\subfigure[Overlayed $3\sigma$ contours.]{		
		\label{subfig:EBL-Overlayed3SigmaContours}
		\includegraphics[width=3.1in]{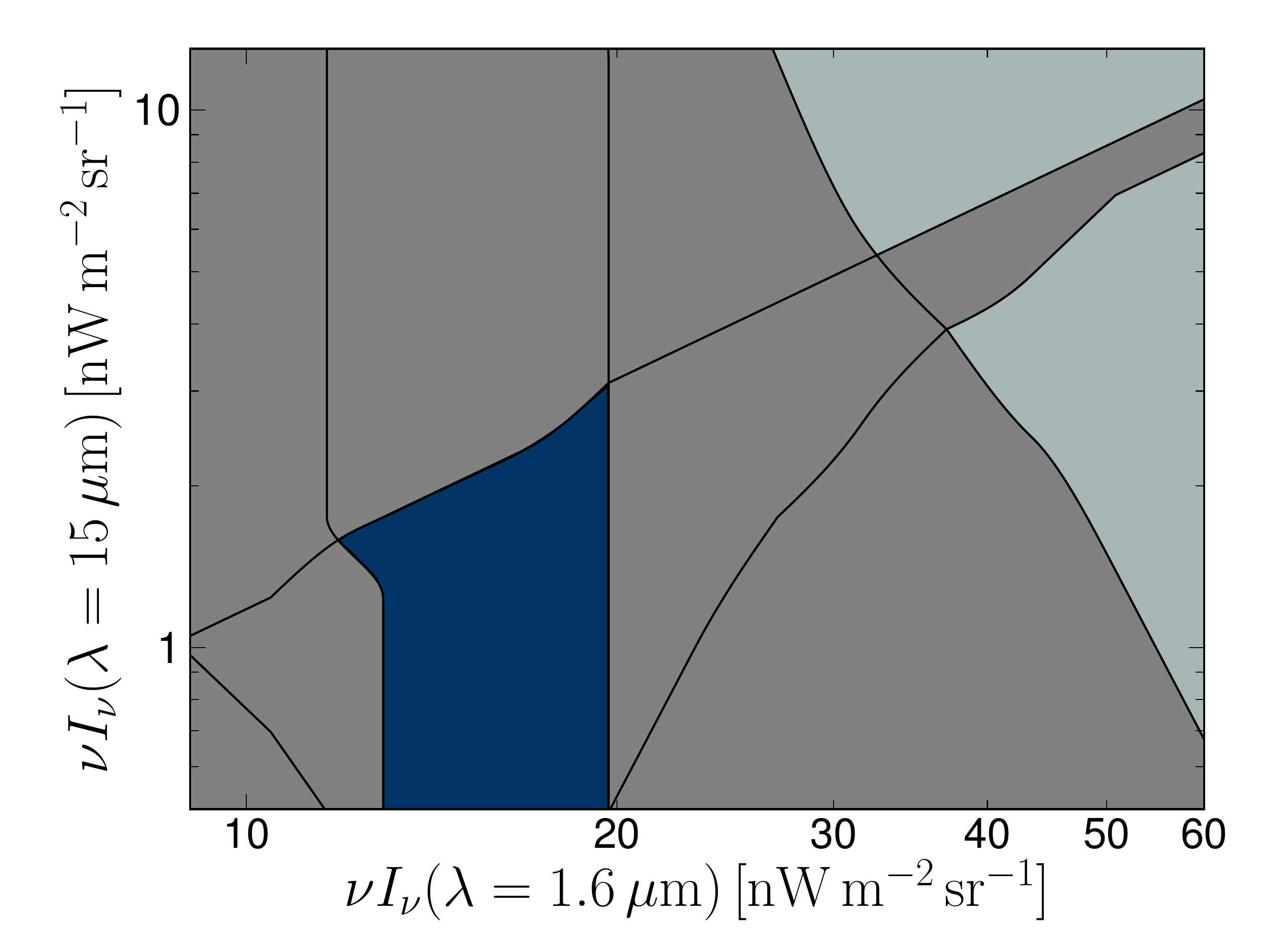}
	}
	\caption{Combined contours from the analysis of RGB J0710+591, 1ES 1218+304, and $\Delta\Gamma(z)$.  The contours of 1ES 1101-232 and 1ES 0229+200 are left off because they do not further constrain the region of allowable parameter space beyond that shown in the figure. The separately colored shaded area in each panel indicates the region of overlap for the three contours.}
	\label{fig:EBL-OverlayedContours}
\end{figure}

\begin{figure}[t]	
	\centering	
	\includegraphics[width=5.25in]{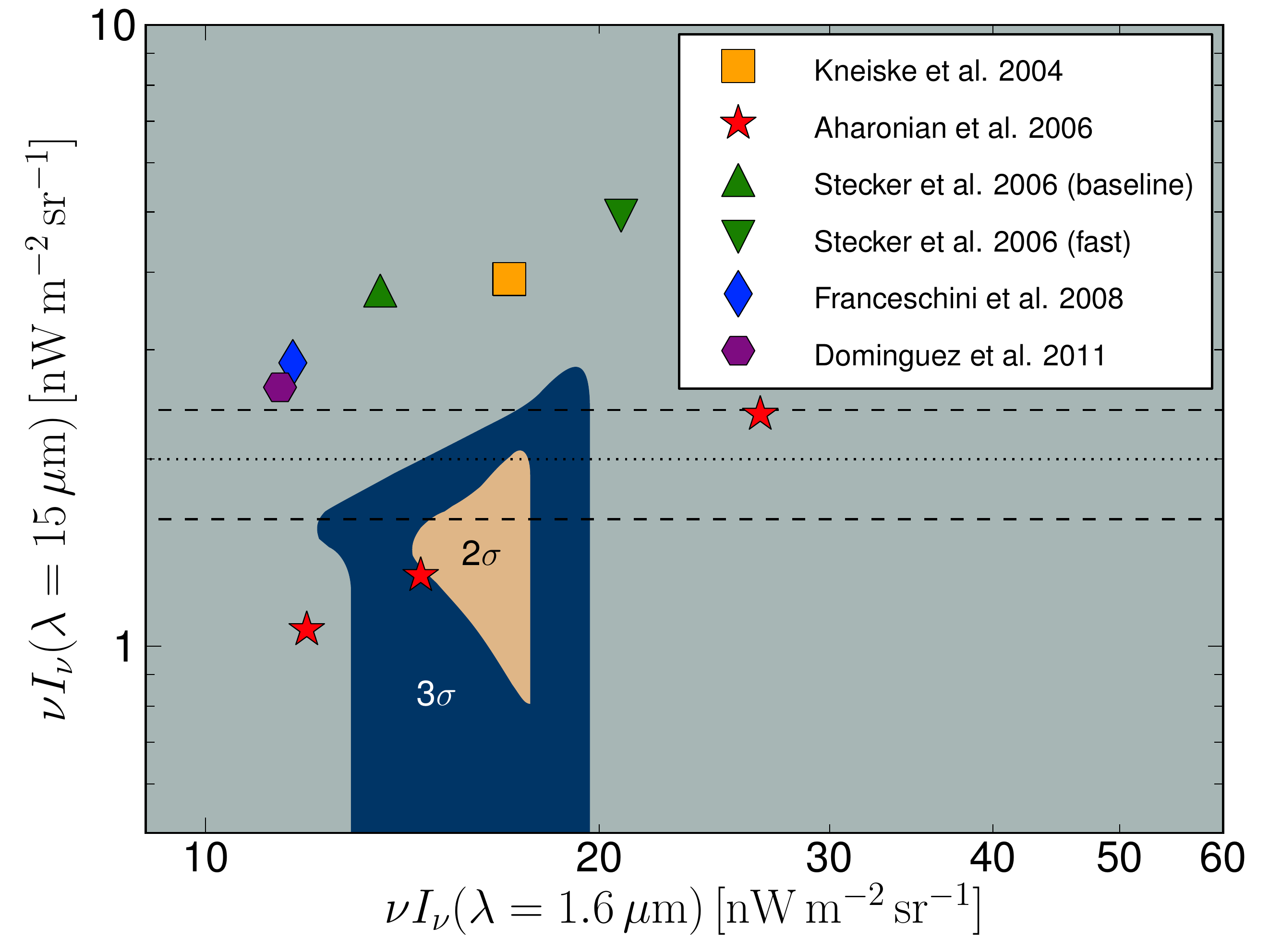}
	\caption{Combined $2\sigma$ and $3\sigma$ contours from the analysis of RGB J0710+591, 1ES 1218+304, and $\Delta\Gamma(z)$.  The contours of 1ES 1101-232 and 1ES 0229+200 are left off because they do not further constrain the region of allowable parameter space beyond that shown in the figure.  Also shown for reference are the models of: \citet{Aharonian:2006} (stars) scaled by 1.0, 0.55, and 0.45 (right to left); \citet{Stecker:2006} (triangles) fast evolution (right) and baseline (left) models;  \citet{Franceschini:2008} (diamond); \citet{Kneiske:2004} (square); \citet{Dominguez:2010} (hexagon).  The dotted line at $\nu I_{\nu}(15 \, \mu\mathrm{m}) = 2.0 \, \mathrm{nW} \, \mathrm{m}^{-2} \, \mathrm{sr}^{-1}$ indicates the total integrated EBL intensity of \citet{Hopwood:2010}. The dashed lines at $\nu I_{\nu}(15 \, \mu\mathrm{m}) = 1.6 \, \mathrm{nW} \, \mathrm{m}^{-2} \, \mathrm{sr}^{-1}$ and $\nu I_{\nu}(15 \, \mu\mathrm{m}) = 2.4 \, \mathrm{nW} \, \mathrm{m}^{-2} \, \mathrm{sr}^{-1}$ indicate the $1\sigma$ error on the value.}
	\label{fig:EBL-CombinedContours}
\end{figure}

Improved constraints on the EBL can be obtained by combining the results of the analyses from Sections \ref{subsec:GeVTeVAnalysis} and \ref{subsec:BreakAnalysis}.  Figure \ref{fig:EBL-OverlayedContours} shows the overlay of $2\sigma$ (a) and $3\sigma$ (b) contours from the analysis of RGB J0710+591, 1ES 1218+304, and $\Delta\Gamma(z)$. The contours from 1ES 1101-232 and 1ES 0229+200 are left off because they do not further restrict the constraints.  Figure \ref{fig:EBL-CombinedContours} shows the final allowed $2\sigma$ and $3\sigma$ regions of parameter space.  The overlapping region indicates a very low EBL intensity at $15\,\mu$m.  It should be reiterated that the assumption here was that the intrinsic TeV  spectrum was approximately equal to the \textit{Fermi} spectrum.  If, in fact, there is an intrinsic break between the GeV and TeV spectra of blazars, the intrinsic TeV spectrum will be softer than is assumed here.  This will move the \textit{Fermi}-IACT derived EBL contours toward the left of Figure \ref{fig:EBL-CombinedContours} and hence allow for lower EBL intensities at $1.6\,\mu$m.  However, given the $\Delta\Gamma(z)$ derived contour (Figure \ref{fig:EBL-TeVBreakContours}), the EBL intensity at $15\,\mu$m will then be constrained to even lower values. 

\begin{figure}[t]	
	\centering	
	\includegraphics[width=5.25in]{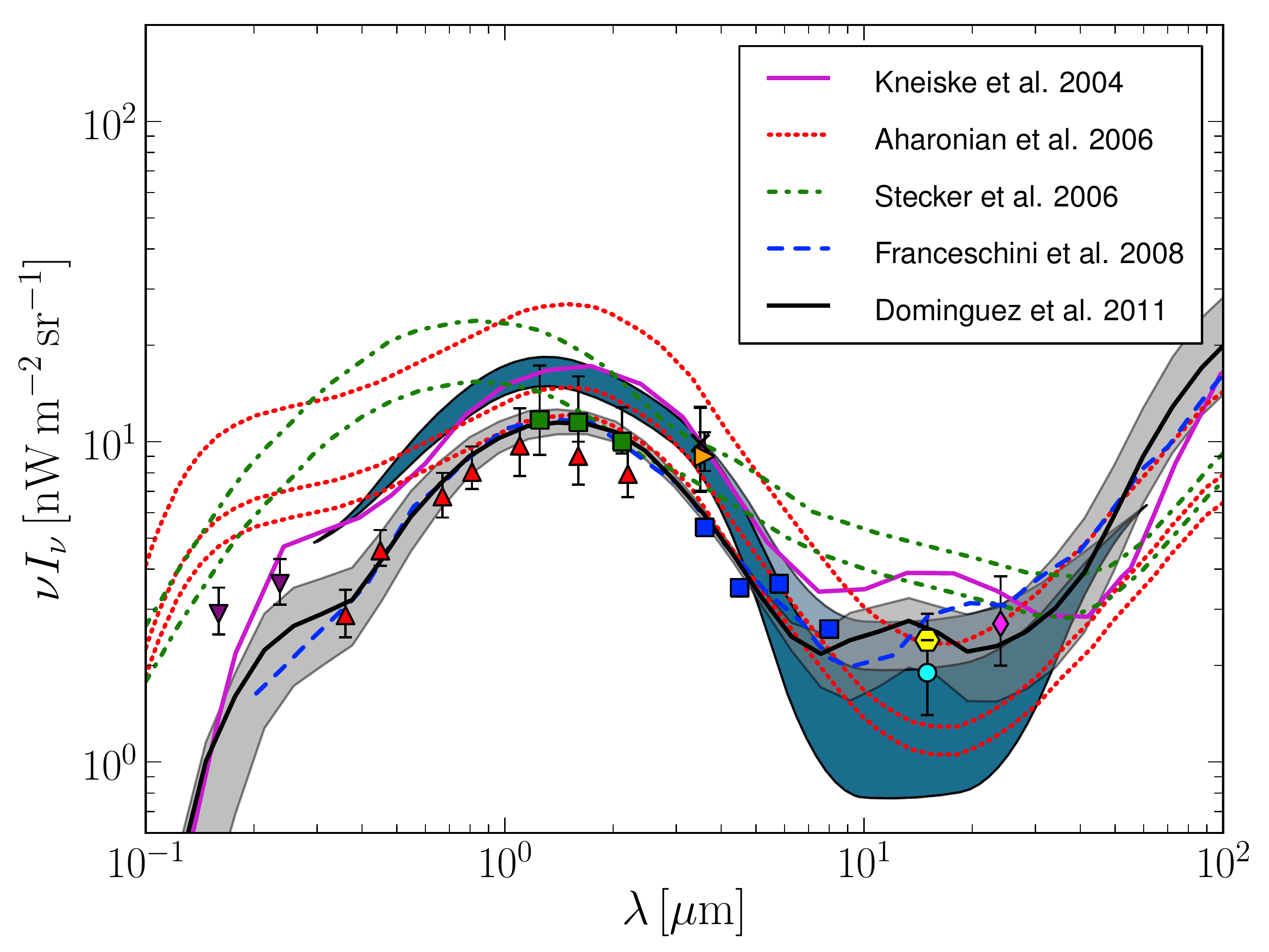}
	\caption{Approximate EBL SED $2\sigma$ confidence region (blue shaded area) derived from Figure \ref{fig:EBL-CombinedContours}.  Also shown are the models of: \citet{Aharonian:2006} (dotted lines) scaled by 1.0, 0.55, and 0.45; \citet{Stecker:2006} (dot-dashed lines) fast evolution (upper) and baseline (lower) models;  \citet{Franceschini:2008} (dashed line); \citet{Kneiske:2004} (upper solid line); \citet{Dominguez:2010} (lower solid line and grey shaded area). The filled symbols indicate the galaxy count lower limits of (left to right): \citet{Gardner:2000} (downward triangles); \citet{Madau:2000} (upward triangles); \citet{Keenan:2010} (squares with error bars); \citet{Levenson:2008} (sideways triangle); \citet{Fazio:2004} (squares without error bars); \citet{Elbaz:2002} (hexagon); \citet{Hopwood:2010} (circle); \citet{Papovich:2004} (diamond).  The $\times$ at $3.5\,\mu$m indicates the tentative detection of \citet{Dwek:1998}.  Note: The partially transparent blue shaded area in the mid-IR shows the expansion of the $2\sigma$ confidence region when using the alternate test spectrum described in Section \ref{subsec:SpecBreakCaveats}.}
	\label{fig:EBL-SEDContours}
\end{figure}

Figure \ref{fig:EBL-SEDContours} shows the $2\sigma$ confidence region for the EBL SED.  The models of \citet{Aharonian:2006}, \citet{Stecker:2006}, \citet{Franceschini:2008}, and \citet{Kneiske:2004} are also shown.  This region is generated using the rectangular area encompassing the $2\sigma$ contour in Figure \ref{fig:EBL-CombinedContours}. 

\section{Caveats and Considerations Regarding Methods 1 \& 2}
\label{sec:Caveats}
There are a number of caveats pertinent to the analyses presented here.  While each of these cannot be discussed exhaustively, we address the most relevant topics below.  First, however, we make a few general remarks.

The EBL attenuation factors calculated throughout the paper were done so using the central energy (in logarithmic space) of each spectral energy bin.  Another way of doing this would be to calculate the average attenuation across the entire energy bin.  We have verified numerically that these two approaches result in absorption factors within $\sim \negthickspace 1\%$ of one another given the typical energy bin widths encountered here.

Not every possible shape of the SED was explored with the EBL parametrization used here.  The aim of this work is to constrain the overall shape of the EBL and not to test for various bumps and wiggles.  The reader is therefore cautioned not to view any \textit{individual} data point, derived from galaxy counts, in Figure \ref{fig:EBL-SEDContours} as being inconsistent with our results if it lies outside the designated contour region.  This analysis does not rule out the possibility that there are, for instance, bumps in the EBL spectrum (e.g., at $\sim \negthickspace 15\,\mu$m).  What is demonstrated by this analysis is that the near- to mid-IR ratio of the EBL appears to be larger than is predicted by many models thereby requiring a relatively high intensity in the near-IR (see Section \ref{sec:DiscussionConclusions}).

\subsection{Hard Spectrum Blazars and TeV IC Peak Energies}
The assumption made in Method 1 was that hard spectrum blazars have IC emission components peaking at TeV energies.  We believe this assumption is well motivated (see Section \ref{subsec:GeVTeVSpec}) but, if it were to prove incorrect, and one instead allowed for some level of intrinsic curvature, the lower limits obtained here in the near-IR would be pushed to lower intensities.  One could also interpret the upper boundary of the $2\sigma$ contour in Figure \ref{fig:EBL-SEDContours} as an upper limit.  However, there is a very important caveat to this second interpretation.  If the near-IR intensity decreases, the mid-IR intensity must also decrease so that the overall slope of the EBL SED between near- and mid-IR wavelengths remains within the allowed range determined with Method 2. 

It should also be noted that the $1\sigma$ contours shown in Figure \ref{fig:EBL-BlazarContours} do not all overlap in a common region of EBL parameter space.  This could be interpreted as being due to intrinsic differences between some of the blazars (e.g., different intrinsic spectral breaks).  However, given that this is a $1\sigma$ effect, and a common overlapping region for the $2\sigma$ contours can be identified, we do not consider this issue to be of significance.

\subsection{Spectral Break Versus Redshift Calculation}
\label{subsec:SpecBreakCaveats}
There are multiple ways of to go about calculating the spectral break versus redshift distribution for a particular EBL scenario.  The method implemented here (Section \ref{subsec:SpecBreakTeV}) used a test blazar spectrum representative of the average spectrum of the 12 blazar data sample.  This determination governed both the energy range covered and the errors on each flux point.  An absorbed spectrum was then calculated for a uniformly spaced set of redshifts.  

An alternative to this approach is to use 12 different test blazar spectra, with each one being representative of a particular blazar in the dataset.  In this case, one would calculate the absorbed spectrum, at each redshift of the observed blazars, using a test spectrum with energy bins and error bars equivalent to the associated observed blazar spectrum.  This technique is most appropriate when testing EBL models that contain features in the SED (e.g., the bump at $\sim \negthickspace 15 \mu$m in the model of \citet{Dominguez:2010}).  In cases such as this, the spectral break calculated, due to EBL absorption, can be sensitive to the energy range over which one performs the fit.  Using the same energy range and binning of the spectral observations allows one to determine the expected spectral break given the specific dataset and an assumed EBL scenario.  This technique is also preferable when the mid-IR EBL intensity is quite high (e.g., the \citet{Stecker:2006} fast evolution model).

We have used the technique described above to compare the various EBL realizations of other authors with the spectral break versus redshift data.  The exclusion probabilities obtained are shown in Table \ref{tab:TeVBreakAnalysisSummary} along side those calculated using the simple best linear fit to the spectral break versus redshift distribution for each scenario.  With the exception of the fast evolution model of \citet{Stecker:2006}, the values obtained using the two techniques are comparable.  The reason for the $1.5 \sigma$ difference in the results of the \citet{Stecker:2006} fast evolution model is likely due to the fact that it has the highest EBL intensity, and therefore the largest amount of absorption, which makes it the most sensitive to the energy range over which the absorbed spectra are fit.  

\begin{figure}[t]	
	\centering	
	\includegraphics[width=5.25in]{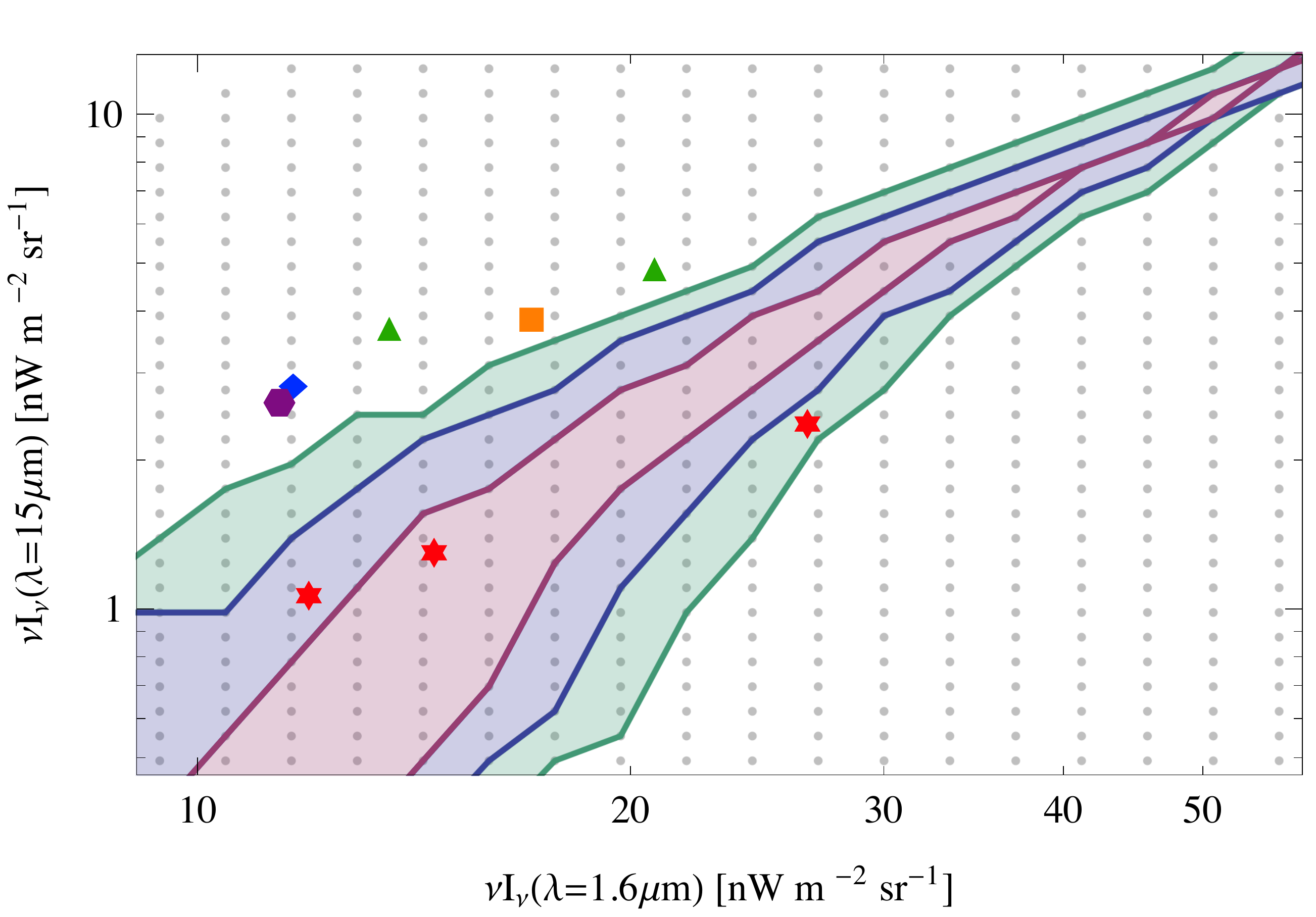}
	\caption{Constraints on the EBL intensity, provided by Method 2, using a test spectrum with 6 spectral points ranging from $400\,$GeV to $3\,$TeV.  See Figure \ref{fig:EBL-TeVBreakContours} for a full description of the contours and data points shown.}
	\label{fig:EBL-TeVBreakContoursAlt}
\end{figure}

As an additional cross check of Method 2, we performed the exact same analysis described in Section \ref{subsec:SpecBreakTeV} using a test blazar spectrum with 6 spectral points ranging in energy from $400\,$GeV to $3\,$TeV.  This more limited energy range around $1\,$TeV elucidates how the low and high energy portions of the spectrum affect the spectral break calculation.  The contours obtained using this test spectrum are shown in Figure \ref{fig:EBL-TeVBreakContoursAlt}.  The main effect of the reduced energy range is to shift the contours upward to slightly higher mid-IR intensities. This is due to the fact that, by excluding energies above $3\,$TeV, the spectral fit is less impacted by the sharp rise in absorption, above a few TeV, that can occur for high mid-IR EBL intensities.  This results in the calculation of a spectral break versus redshift distribution with a more positive, or less negative, slope which allows for higher mid-IR intensities.  The corresponding $2\sigma$ contour region for the EBL SED is shown in Figure \ref{fig:EBL-SEDContours} as the partially transparent blue shaded region.

It is also important to understand the effects of each individual source on the observed spectral break versus redshift distribution shown in Figure \ref{fig:TeVBreakVsZ}.  Namely, how do the results change if sources lying $\gtrsim \negthickspace 2\sigma$ outside the fit (e.g., PKS 0548-322 and H 1426+428) are removed?  By excluding H 1426+428 from the dataset, the fit result of Figure \ref{fig:TeVBreakVsZ} changes by only $\sim \negthickspace 5\%$.  Removing PKS 0548-322 from the dataset results in a best fit of $\Gamma (z) = (12.68 \pm 5.79) z - (0.85 \pm 0.78)$.  In this case, both the slope and intercept of the fit change by less than $1 \sigma$ while the exclusion significances of the various EBL models mentioned in the paper (e.g., \citet{Primack:2005}, \cite{Stecker:2006}, etc.) change only marginally.

\subsection{Evolution of the EBL with Redshift}
The evolution of the EBL as a function of redshift should be taken into account when calculating the gamma-ray absorption for sources with large redshifts.  Since the quality of the data used, and the nature of the analysis performed, will impact exactly at what redshift EBL evolution becomes a necessary inclusion, we investigate its effects here.  

The EBL is comprised of photons generated by sources that have been evolving since the early Universe.  This evolution is, therefore, imprinted on the EBL but has been neglected in  our analysis thus far.  This means that, when calculating the absorption for a source at a redshift of, say, $z=0.2$, we have been assuming that all of the target EBL photons are in place at this redshift and that there is no absorption and/or reemission of these photons (i.e., no evolution) between $z=0.2$ and $z=0$.  This assumption is reasonable for small redshifts, but one must determine at what redshift it is no longer warranted.  

The evolution of the EBL has been examined in detail by many of the authors whose models are used here (e.g., \citet{Kneiske:2004} and \citet{Franceschini:2008}).  There is also a simple first order approach to incorporating EBL evolution into absorption calculations.  This method simply applies a correction factor to the cosmological photon numbers density scaling (e.g., \citet{Madau:1996} and \citet{Raue:2008}).  This factor is chosen such that the resultant EBL evolution is in good agreement with more detailed models (e.g., \citet{Kneiske:2002}).  \citet{Raue:2008} have shown that this approach can be used to match the EBL evolution well for redshifts of $ z \lesssim 0.7$.

Using this approach, we have investigated the effects of EBL evolution on the results presented here.  The inclusion of EBL evolution lowers the calculated optical depths by $\sim \negthickspace 10\%$ at a redshift of $z=0.2$.  For the blazars analyzed in Method 1, this translates to a softening of the intrinsic spectra by $\sim \negthickspace 0.5 \sigma$.  This corresponds to a shift of $\sim \negthickspace 10\%$, to higher EBL intensities, in the contours of Figure \ref{fig:EBL-BlazarContours}.

Method 2 is insensitive to the intrinsic spectral softening effects due to the incorporation of EBL evolution since this method relies on the measurement of the spectral break and not the spectral index itself.  In other words, it depends on a relative measurement rather than an absolute one.  It should be noted, however, that this is contingent upon the (reasonable) assumption that only the intensity of the EBL evolves over such small redshifts and not the shape of the SED.  In other words, the relative contributions to the EBL from different emission components remain approximately the same over small redshifts. 

\subsection{Alternatives Scenarios Describing Blazar Spectral Observations}
Alternative scenarios that could explain the observed spectra of blazars, while allowing for a wider range of EBL intensities, have been investigated by others.  Some have proposed that the observed high energy gamma-ray emission from blazars could be dominated by secondary photon emission produced along the line of site by cosmic-ray interactions with EBL photons \citep{Essey:2010:AstroPart,Essey:2010:arXiv,Essey:2010:PhysRev}.  \citet{Essey:2010:arXiv} have shown that such a scenario can be used to describe the observed spectra of many VHE blazars.  Other authors have suggested that photons could oscillate into light axion-like particles (ALPs) via interactions with intergalactic magnetic fields \citep{DeAngelis:2007,DeAngelis:2009}.  These ALPs, which travel unimpeded through the Universe, can then convert back into photons before reaching the Earth.  Since ALPs are not affected by EBL absorption, this process would allow for an EBL intensity that is, in actuality, higher than would appear based on standard absorption calculations.  These are just two of the many interesting avenues of research being pursued in relation to the EBL.

\section{Improving Constraints on the EBL}
\label{sec:ImprovingConstraints}
The constraints on the EBL presented here can be improved through further observations with \textit{Fermi} and IACTs.  \textit{Fermi} is continuously monitoring the entire sky and hence data on the sources used in Method 1 are being acquired all the time.  Improved measurements in the TeV regime using current instruments will require either deep or flaring state observations of blazars and will predominantly impact the constraints obtained with Method 2.  As can be seen from Figure \ref{fig:TeVBreakVsZ}, the majority of the blazar spectral breaks measured using current data are not, on an individual basis, statistically significant.  The driving factor behind the large measurement errors is an insufficient knowledge of blazar emission at energies $> \negthickspace 1\,$TeV for most of the distant sources.  As deeper exposures are obtained, and source photon statistics continue to improve, measurements of many blazars will extend to ever higher energies until their spectra either intrinsically, or via EBL absorption, cut off. 

As spectra span a larger energy range, and as the precision of each flux point improves, it is less likely that their intrinsic emission will be well described by a single power-law.  When looking for evidence of a break at $\sim \negthickspace 1\,$TeV, it would be necessary to modify the approach used here by either truncating the fit regime so as to exclude energies above or below where the IC peak is believed to be located, or use a more complicated fit function that includes (at least) a spectral break parameter along with a curvature parameter.  It will also likely be necessary to use the more sophisticated spectral break versus redshift determination described in Section \ref{subsec:SpecBreakCaveats}.

Another consideration, regarding the detection of blazars in flaring states, is that of spectral variability.  Some blazars exhibit spectral hardening with increasing flux \citep{Krennrich:2002,Albert:2007:Mrk421,Albert:2007:Mrk501} while others show no evidence for spectral variability when flaring \citep{Acciari:2010:1ES1218}.  Any change in the overall slope of the spectrum would have a limited impact on Method 2, given that it relies on the relative change in spectral index (spectral break), but the limits obtained using the \textit{Fermi} and IACT data as in Method 1 could be affected.\footnote{The four sources used in Method 1 show no signs of spectral variability.}  If the TeV spectrum were to harden, while the \textit{Fermi} spectrum remained unchanged, tighter constraints on the EBL could be obtained due to the smaller observed spectral break.  However, if spectral variability takes place entirely above or below the IC peak, for hard spectrum blazars, where we have assumed that the IC peak is at multi-TeV energies, both \textit{Fermi} and IACTs should see the same amount of spectral hardening thereby having no impact on the derived constraints.  Simultaneous \textit{Fermi} and IACT data are needed during both flaring and quiescent states to truly evaluate the effect (if it exists) of spectral variability on hard spectrum blazars.

\section{Discussion \& Conclusions}
\label{sec:DiscussionConclusions}
We have presented constraints on the EBL intensity using two distinct methods for the analysis of blazar spectra. The spectral shape method (Method 1) used \textit{Fermi} observations of the three hard spectrum blazars RGB J0710+591, 1ES 1101-232, and 1ES 1218+304 as a proxy for the intrinsic source emission in the sub-TeV to TeV energy regime. In addition, 1ES 0229+200, which has a weak \textit{Fermi} detection ($\sim \negthickspace 4 \sigma$), was used in this analysis, with an assumed \textit{Fermi} spectral index of 1.5, as its elucidates the impact of sources detected above $\sim \negthickspace 10\,$TeV on EBL constraints.  Observed IACT spectra were corrected for attenuation due to different EBL realizations. Limits on the EBL were then placed by rejecting realizations that did not produce deabsorbed (i.e., intrinsic) IACT spectra consistent with the \textit{Fermi} extrapolation.

The second method, the TeV spectral break method, involved two steps. The first consisted of characterizing the spectral break, occurring at $\sim \negthickspace 1\,$TeV, using the IACT observations of 12 blazars located at different redshifts. The magnitude of the spectral break was then plotted as a function of redshift (Figure \ref{fig:TeVBreakVsZ}). We then assumed that the intrinsic blazar spectrum can be characterized as a power-law which, unlike Method 1, was not determined by the \textit{Fermi} spectrum, and is therefore more general in nature. The second step consisted of comparing the data in Figure \ref{fig:TeVBreakVsZ} with the redshift distribution of spectral breaks derived from the attenuation of an intrinsic test spectrum using different EBL realizations. 

We showed that Method 1 primarily constrains the overall EBL intensity, whereas Method 2 primarily constrains the relative EBL intensities in the near- and mid-IR ($\sim \negthickspace 1.6 \, \mu$m and $\sim \negthickspace 15 \, \mu$m, respectively). The two approaches are therefore complementary and provide a closed confidence range for the EBL intensity. This is the first time these methods have been used to derive such a contour for the EBL SED. 

The resulting constraints imposed by both methods on the the $1.6 \, \mu$m and $15 \, \mu$m intensity of the EBL are shown in Figure \ref{fig:EBL-CombinedContours}. Using ultra deep $15 \, \mu$m mapping observations of the gravitational lensing cluster Abell 2218 with the AKARI infrared satellite, \citet{Hopwood:2010} derived an intensity of $1.9 \pm 0.5 \, \mathrm{nW} \, \mathrm{m}^{-2} \, \mathrm{sr}^{-1}$ for the integrated galaxy light (IGL) down to an intensity of $\sim \negthickspace 0.01\,$mJy. A model fit to the differential contribution to the IGL, combining data from several fields, gives a total integrated EBL intensity of $2.0 \pm 0.4 \, \mathrm{nW} \, \mathrm{m}^{-2} \, \mathrm{sr}^{-1}$ \citep{Hopwood:2010}. The IGL seems to have totally resolved the EBL at $15 \, \mu$m. Using our constraints, and the claimed $15 \, \mu$m EBL detection, we derive a value of $17 \pm 3 \, \mathrm{nW} \, \mathrm{m}^{-2} \, \mathrm{sr}^{-1}$ for the EBL intensity at $1.6 \, \mu$m, the error representing a $1\sigma$  uncertainty. Comparison with the IGL \citep{Madau:2000} suggests that $\sim \negthickspace 50\%$, and perhaps as much as $\sim \negthickspace 90\%$ \citep{Keenan:2010}, of the EBL has been resolved at this wavelength. 

Figure \ref{fig:EBL-SEDContours} presents the $2\sigma$ confidence range of the EBL intensity derived in this paper as a function of wavelength (shaded blue). Our derived constraints are consistent with claimed EBL detections at $3.5 \, \mu$m \citep{Dwek:1998} and $3.6 \, \mu$m \citep{Levenson:2008}. Figure \ref{fig:EBL-SEDContours} also shows that our EBL constraints are consistent with some models, such as the scaled version of \citet{Aharonian:2006} and the most recent model of \citet{Hopwood:2010}. However, the models of \citet{Stecker:2006}, \citet{Franceschini:2008}, \citet{Kneiske:2004}, and \citet{Dominguez:2010} are disfavored by our analysis at $> \negthickspace 3\sigma$ confidence level. 

By inspection of Figure \ref{fig:EBL-SEDContours}, one may be tempted to conclude that the models of \citet{Franceschini:2008} and \citet{Dominguez:2010} are inconsistent with the analysis presented here predominantly because their intensities in the near-IR are too low.  One might then suspect that this is purely due to constraints placed on the EBL by Method 1, which would be lower if the assumption of a single power-law between GeV and TeV energies was not made.  This, however, would be an incorrect conclusion.  The exclusion probabilities calculated in Table \ref{tab:TeVBreakAnalysisSummary} are entirely independent of Method 1, using only the spectral break versus redshift distribution of Method 2.  As such, what is truly driving the agreement or disagreement of a particular model with the data in this case is the near- to mid-IR ratio.  

The minimum average slope between 2 and $10 \, \mu$m, falling within our $2\sigma$ contour in Figure \ref{fig:EBL-SEDContours}, is $\alpha \negthinspace \sim \negthinspace 1.36$ ($\propto \negthinspace \lambda^{-\alpha}$).  This value is in agreement with the limit placed by \citet{Aharonian:2007} of $\alpha \negthinspace \gtrsim 1.1 \pm 0.25$ using 1ES 0229+200 data from H.E.S.S.  \citet{Aharonian:2007} also calculate an upper limit on the EBL intensity at $10 \, \mu$m of $3.1 \, \mathrm{nW} \, \mathrm{m}^{-2} \, \mathrm{sr}^{-1}$.  This constraint is also compatible with our results, which places an upper bound at $10 \, \mu$m of $\sim \negthinspace 2 \, \mathrm{nW} \, \mathrm{m}^{-2} \, \mathrm{sr}^{-1}$.

In summary, our analysis has shown the advantages of using these two complementary approaches in constraining the EBL.  It has also provided evidence that the near- to mid-IR ratio of the EBL may be larger than previously thought. Future measurements with \textit{Fermi}, as well as both current and next generation IACTs, such as the Cherenkov Telescope Array (CTA), will continue to improve these constraints.

\acknowledgements
FK acknowledges support from DOE.  ED and FK acknowledge support from the \textit{Fermi} Cycle-2 Guest Investigator program.  The authors would also like to thank the anonymous referee for a thorough review of this manuscript which has improved both its clarity and completeness as well as revealed many subtleties regarding the analysis methods developed here. 

\bibliographystyle{apj}
\bibliography{apj-jour,bibliography}

\appendix

\section{Investigation of Systematics}
To obtain a more thorough understanding of how the choice of certain parameter values (flux point error bars, number of iterations, redshift binning, etc.) affect this analysis, we have investigated the potential systematic effects introduced by each.  

\subsection{Test Blazar Spectrum Flux Point Errors}
\begin{figure}[t]
	\centering
	\subfigure[Slope parameter.]{	
		\includegraphics[width=2.775in]{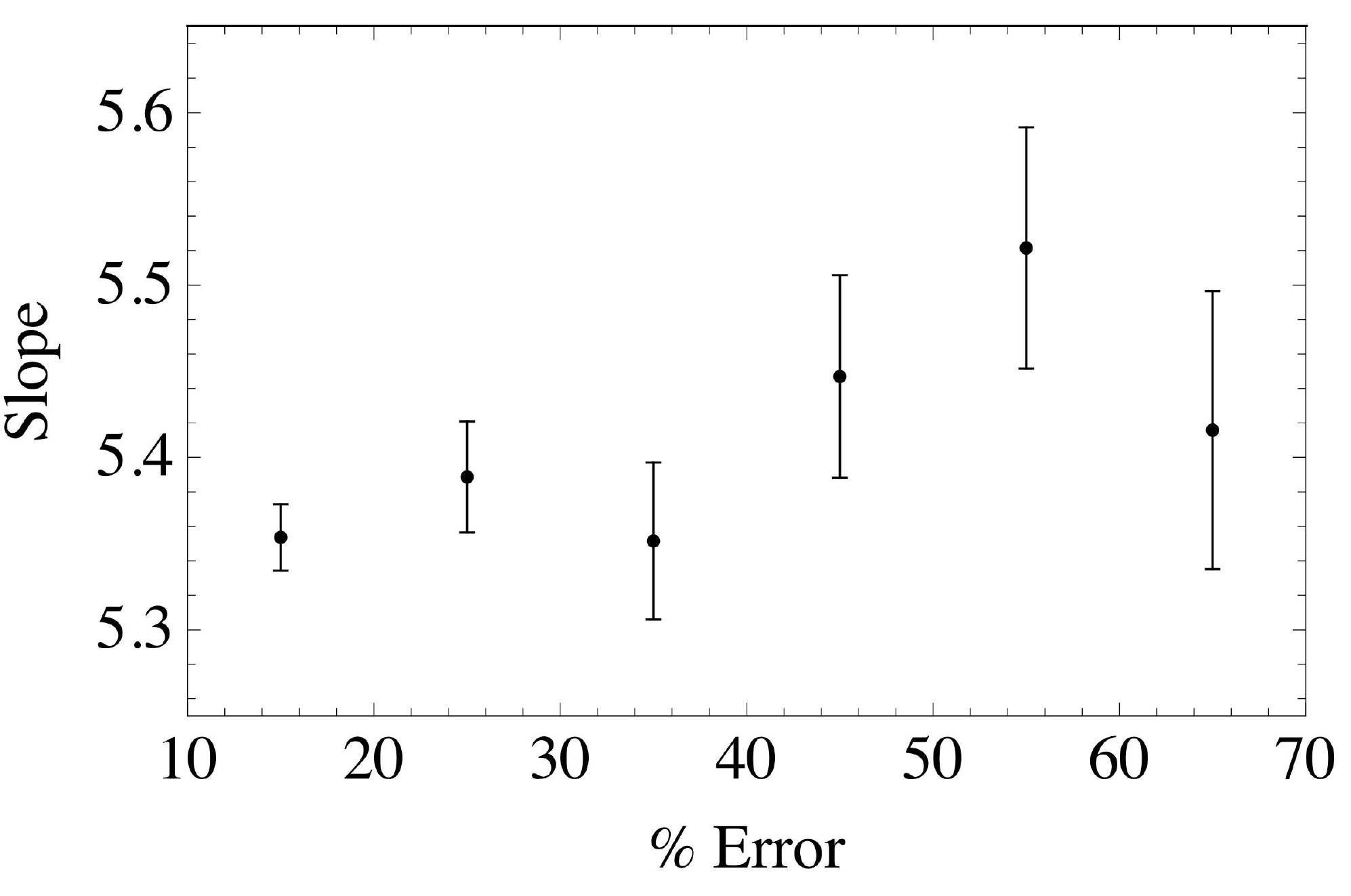}
	}
	\subfigure[Intercept parameter.]{		
		\includegraphics[width=3.1in]{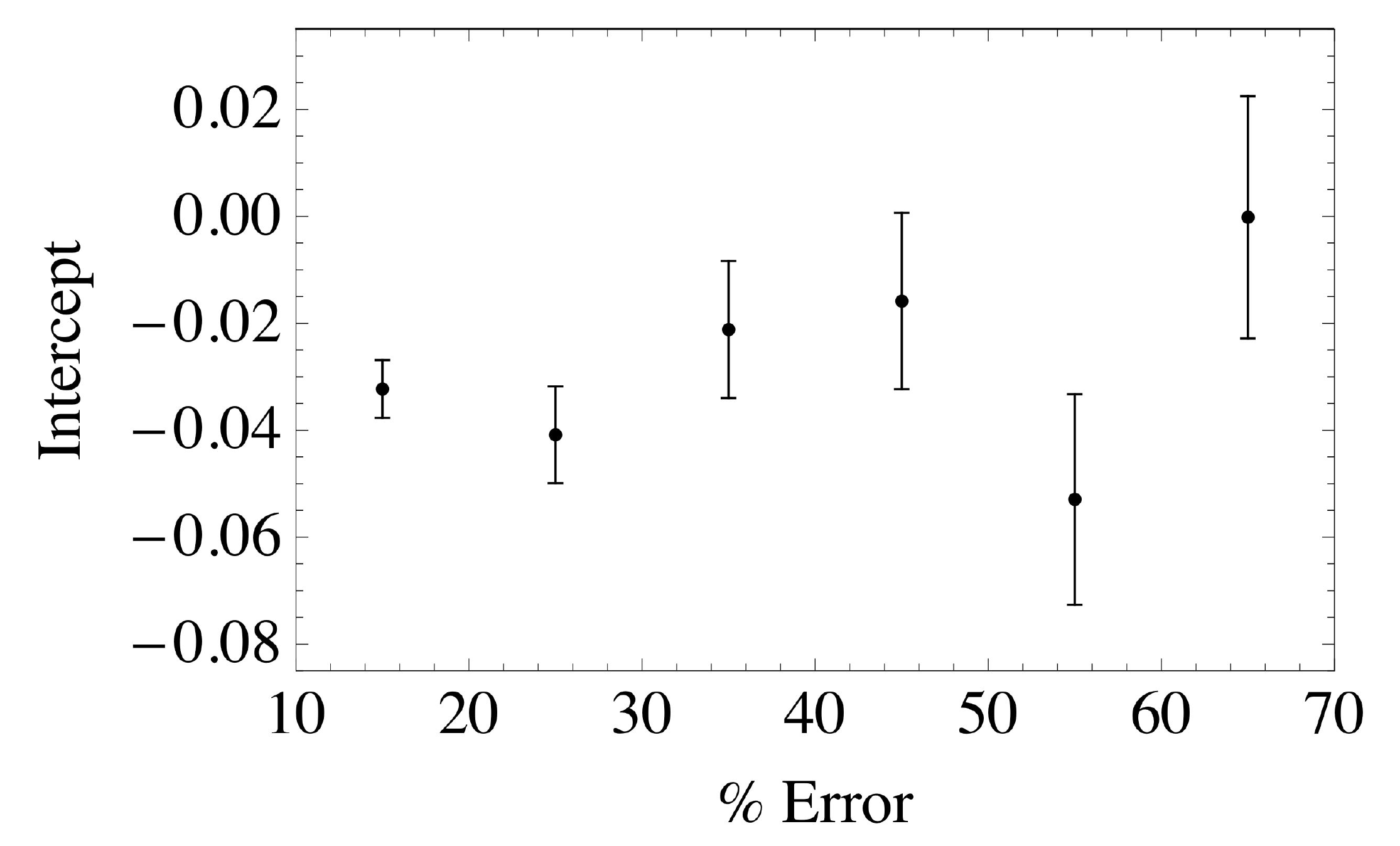}
	}	
	\caption{Figures showing the slope (left) and intercept (right) fit results for the TeV spectral break versus redshift distribution (Equation \ref{eqn:BreakVsZFit}) assuming a particular EBL model and using different size error bars for the test blazar spectrum (see Section \ref{subsec:SpecBreakTeV}).  The six different choices of errors shown are 15\%, 25\%, 35\%, 45\%, 55\%, and 65\%.  The standard deviations (unweighted) of the slope and intercept best fit parameters are 0.065 and 0.020, respectively.  This spread in values is $\lesssim \negthickspace 3\%$ of the error in the best fit determination from observations (Figure \ref{fig:TeVBreakVsZ}).}
	\label{fig:FitParamSystematics}	
\end{figure}

The application of 25\% error bars to the test blazar spectrum (Section \ref{subsec:SpecBreakTeV}), while motivated by the mean percent error from VERITAS spectra, is a somewhat arbitrary choice.  However, the systematic affect on the final result is limited, as illustrated in Figure \ref{fig:FitParamSystematics}.  This figure shows the fit results for the TeV spectral break versus redshift distribution (Equation \ref{eqn:BreakVsZFit}), after 100 iterations, assuming a particular EBL model.  The left (right) panel shows the best fit slope (intercept) using six different size error bars, ranging from 15\%--65\%, for the test blazar spectrum.  While the error in the best fit value increases with the error in the test blazar spectrum,\footnote{This could be remedied by using a larger number of iterations (i.e., simulated observations).} the standard deviation in the final result is $\lesssim \negthickspace 3\%$ of the error in the best fit determination from observations (Figure \ref{fig:TeVBreakVsZ}).  From this one can be conclude that the systematic effect introduced by the choice of error bars in the test spectrum is minimal.  

\subsection{Redshift Binning and the Number of Iterations for $\Delta\Gamma(z)$ Determination}
The choice of binning in redshift and the number of iterations used for the determination of $\Delta\Gamma(z)$ are two potential sources of systematics.  A redshift binning that is too course may result in a poor characterization of the spectral break versus redshift distribution.  Many of the blazars used in the TeV spectral break analysis are separated in redshift from another source in the sample by $\Delta z < 0.05$ (see Table \ref{tab:BlazarSample}).  It may seem then that a redshift binning of 0.05 is insufficient for comparing the $\Delta\Gamma(z)$ distributions for given EBL models with that from observation.  Additionally, the choice of 100 iterations (i.e., simulated observations) for the determination of the $\Delta\Gamma(z)$ distribution is not necessarily sufficient for the fit to converge to its true value.

\begin{figure}[p]
	\centering
	\subfigure[1 iteration.]{	
		\includegraphics[width=3.in]{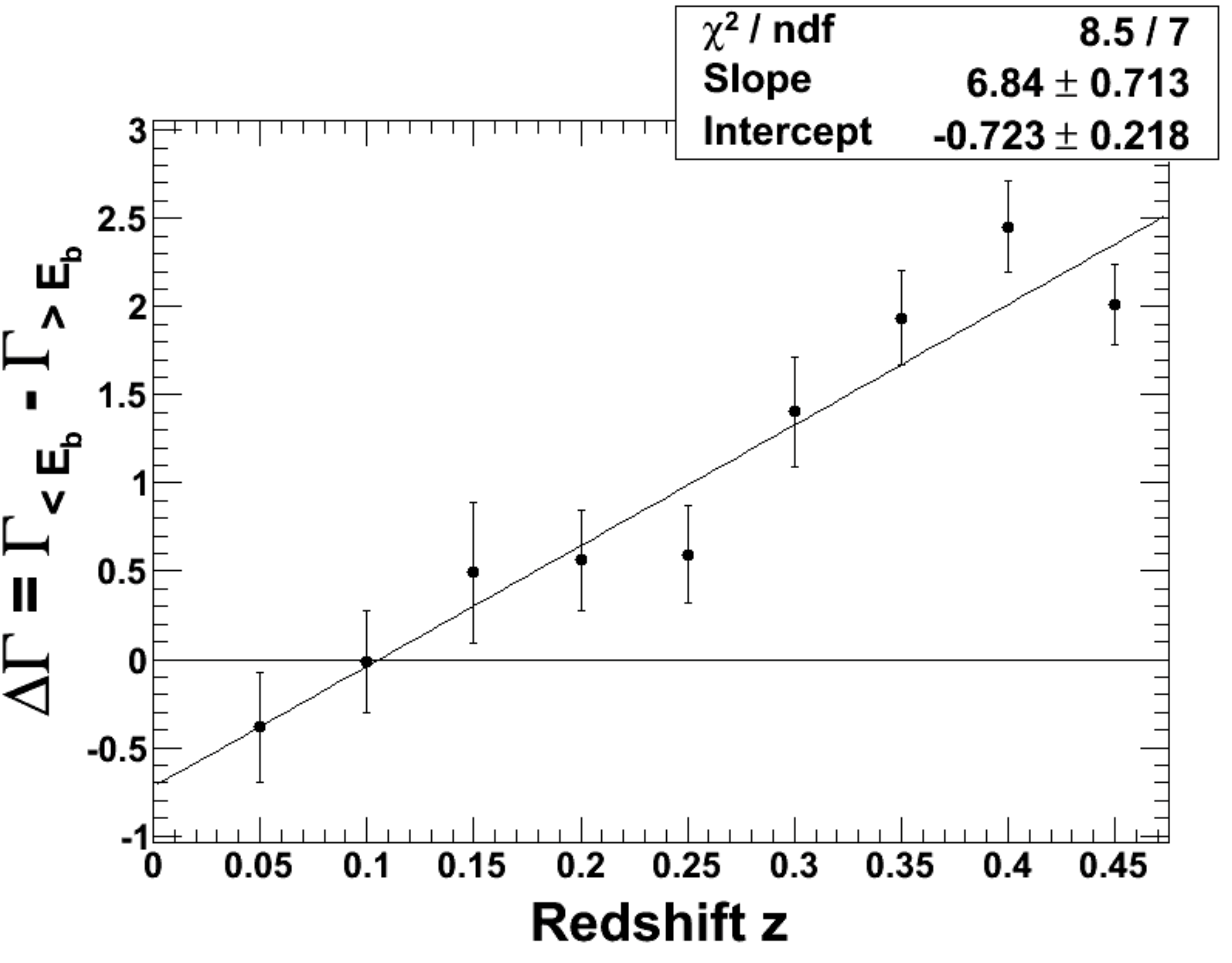}
	}
	\subfigure[25 iterations.]{
		\includegraphics[width=3.in]{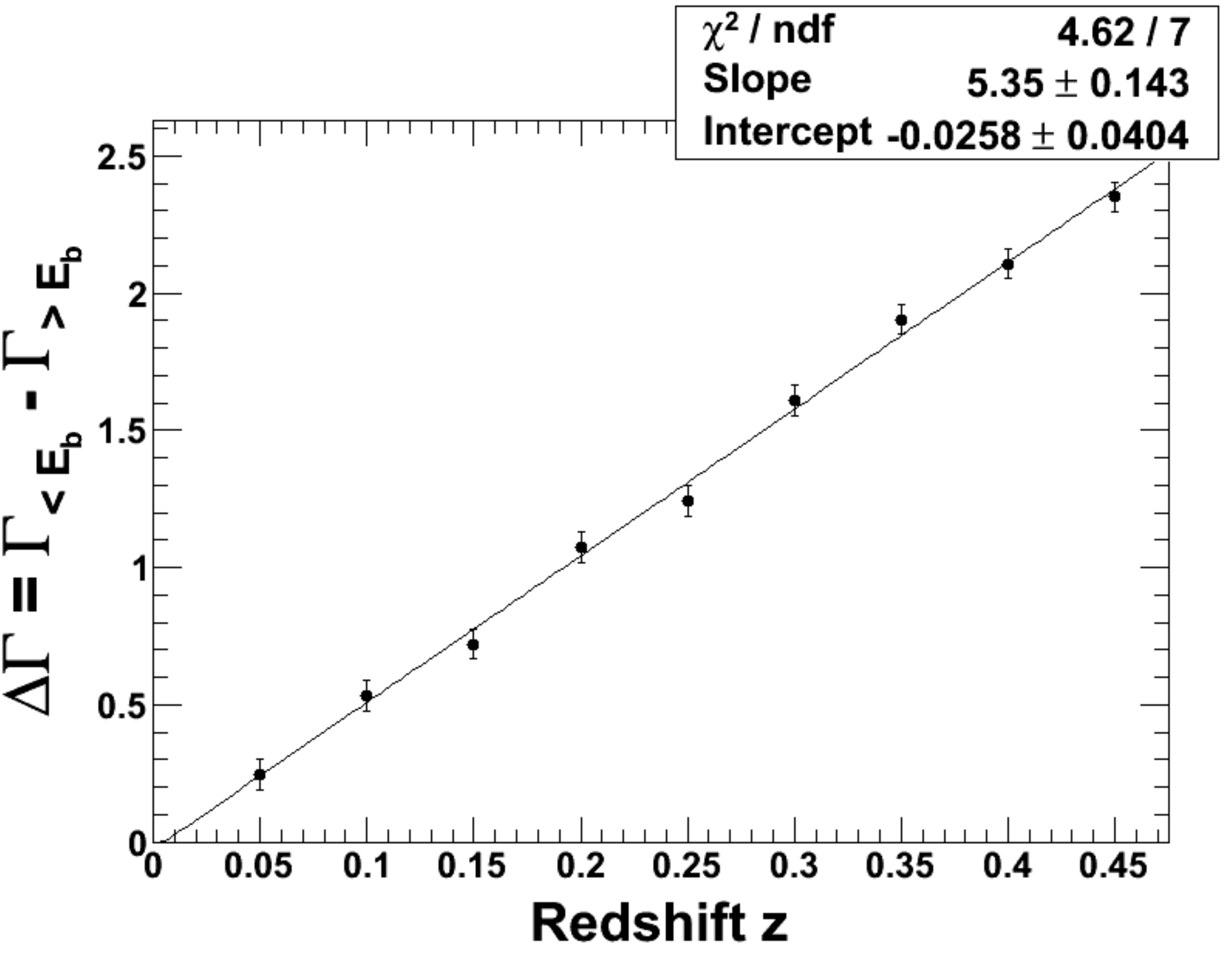}
	}	
	\subfigure[50 iterations.]{	
		\includegraphics[width=3.in]{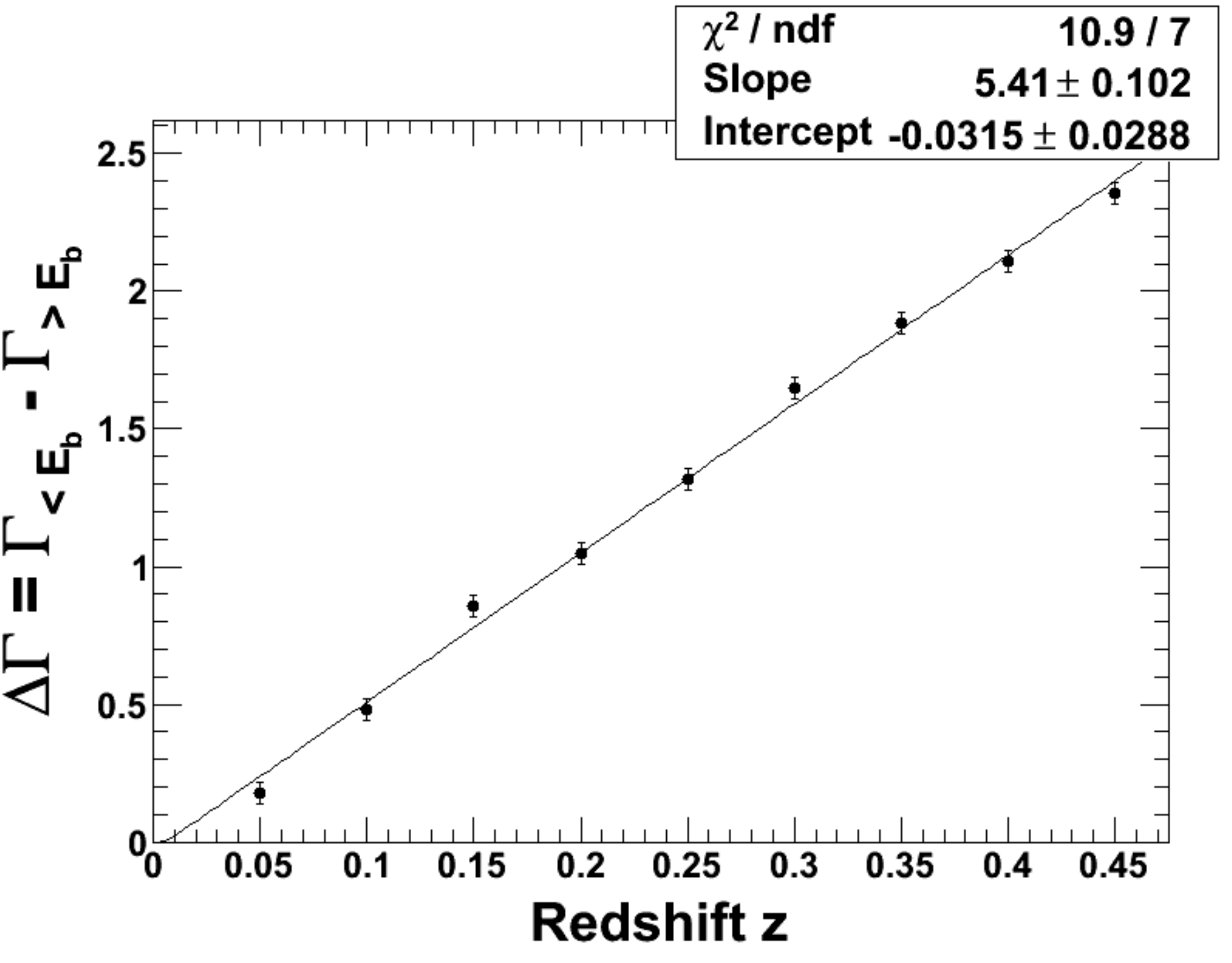}
	}
	\subfigure[100 iterations.]{
		\includegraphics[width=3.in]{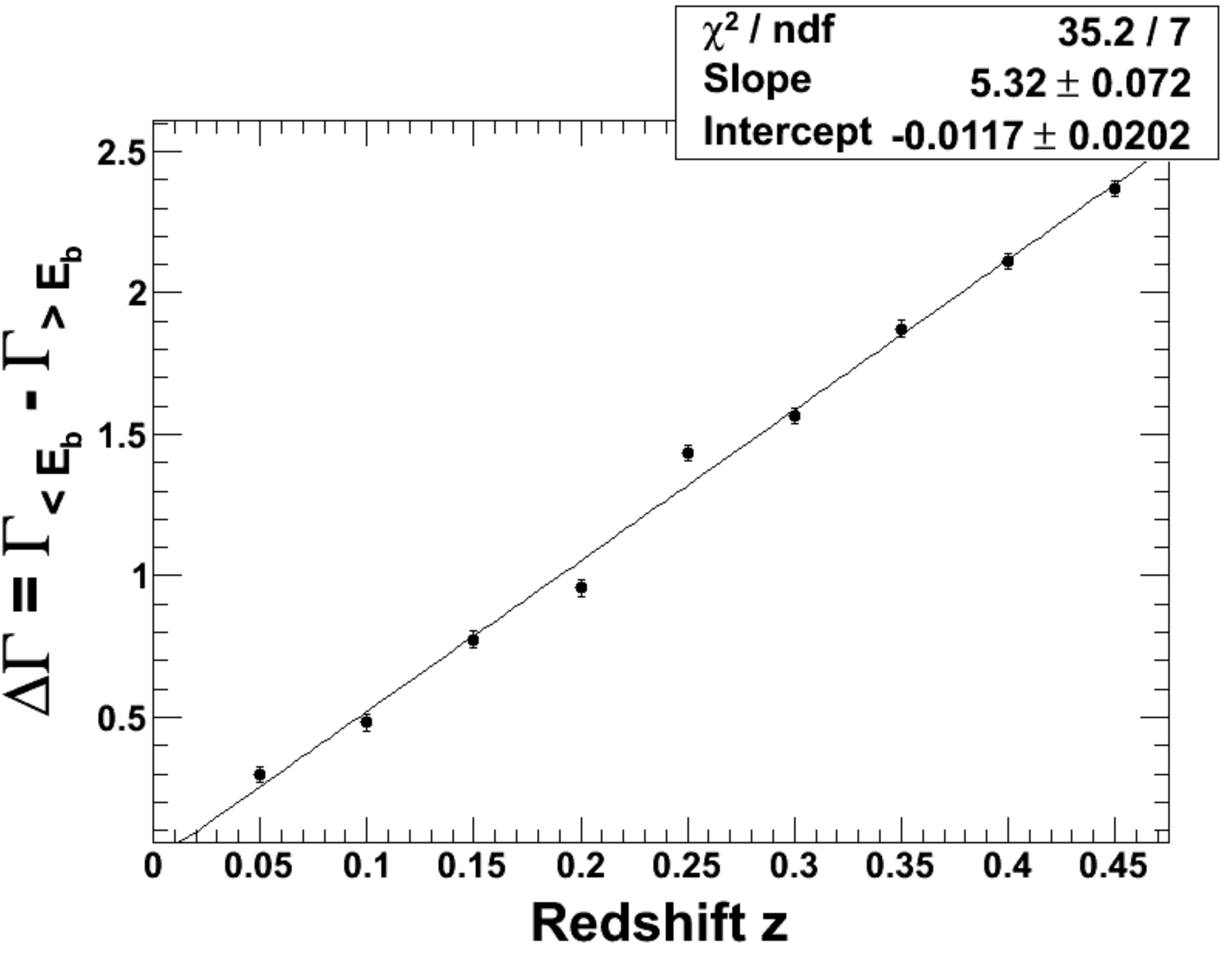}
	}
	\caption{$\Delta\Gamma(z)$ distribution, using EBL absorbed test blazar spectra for a particular EBL scenario, calculated after 1, 25, 50, and 100 iterations.}
	\label{fig:TeVBreakVsZSim}	
\end{figure}

Figure \ref{fig:TeVBreakVsZSim} addresses both these points.  The four panels show the spectral break versus redshift distribution after 1, 25, 50, and 100 iterations using one particular EBL model.  It can clearly be seen that the distribution has converged after 100 iterations.  This is reflected not only in the data point error bars, but in the errors on the best fit parameters as well.  It is also clear that, while the distribution is not perfectly linear, there is no evidence of features from small variations in redshift.  In other words,  a finer binning in redshift would have virtually no impact on the fit results.     

\subsection{Potential Pileup Effect at Very High Energies}
\begin{figure}[t]
	\centering
	\subfigure[]{
		\includegraphics[width=2.9in]{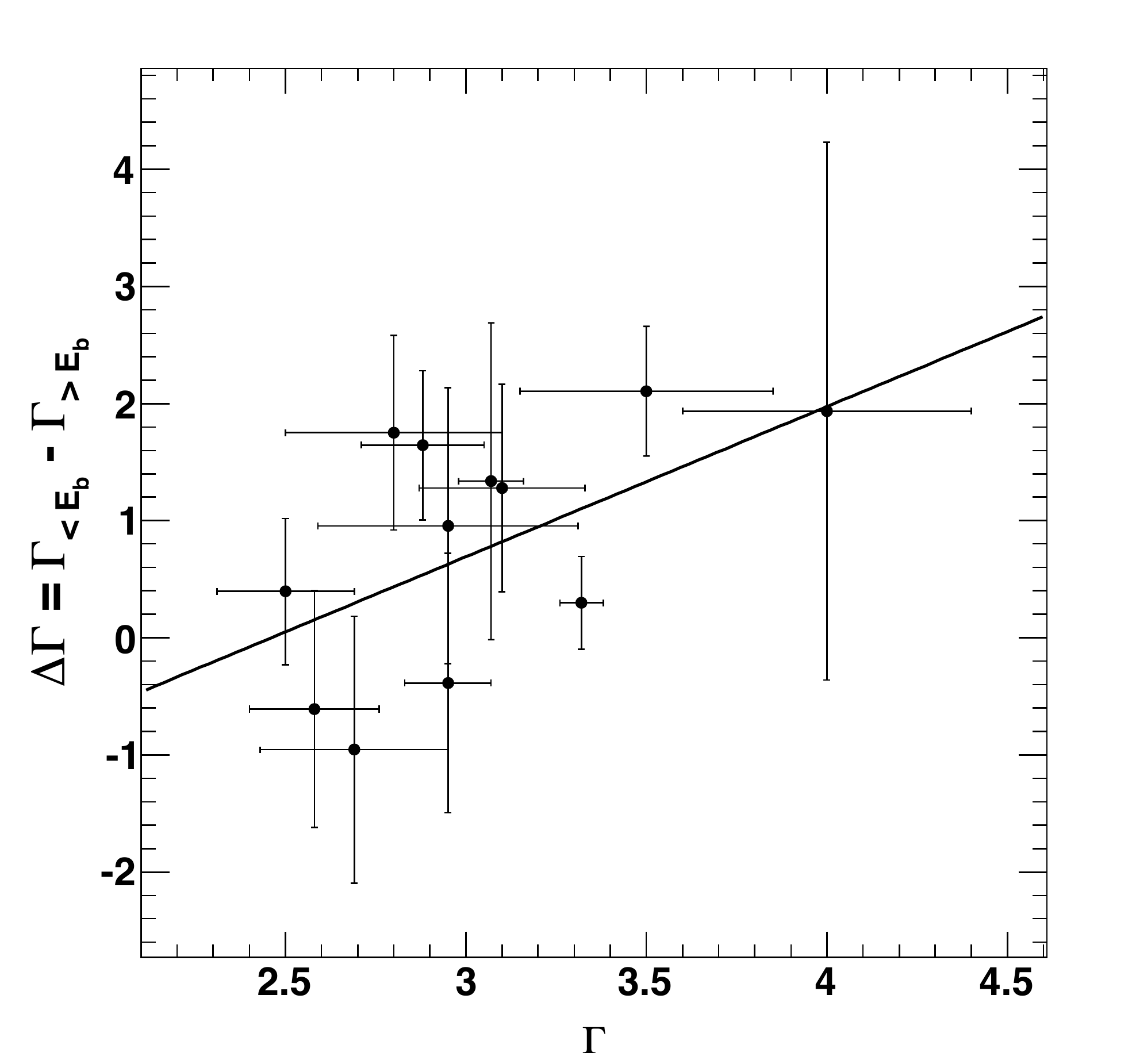}
		\label{subfig:SpecBreakVsGamma}
	}
	\subfigure[]{
		\includegraphics[width=2.9in]{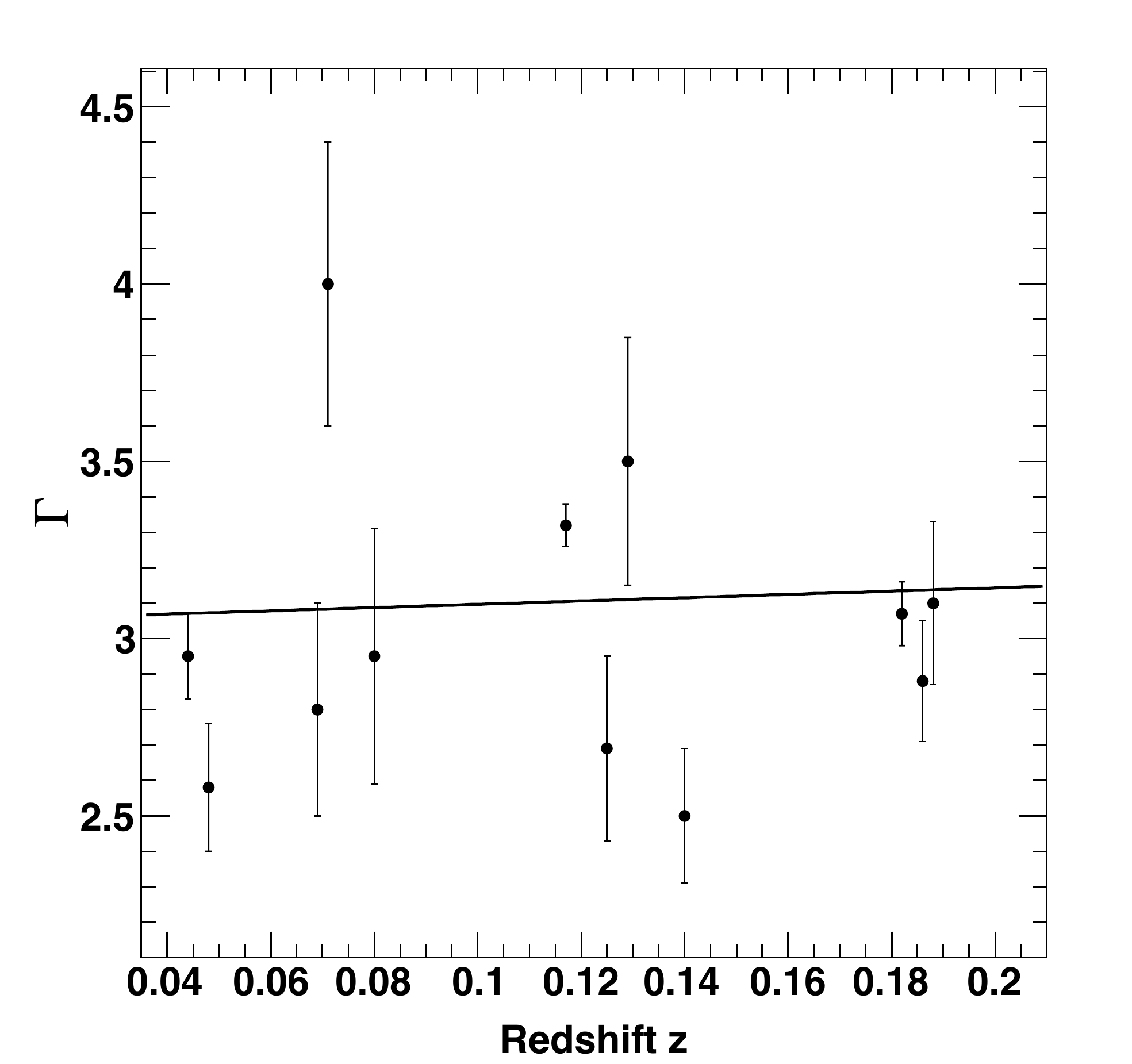}
		\label{subfig:GammaVsZ}
	}
	\caption{(a) Spectral break (as shown in Figure \ref{fig:TeVBreakVsZ}) plotted versus the spectral index obtained when fitting the blazar spectra with a single power-law.  The best linear fit to the data is given by $\Delta\Gamma(\Gamma) = (1.28 \pm 0.68)\Gamma - (3.15 \pm 2.15)$.  (b) Single power-law spectral index plotted versus the blazar redshift.  The best linear fit to the data is given by $\Gamma(z) = (0.46 \pm 0.87)z + (3.05 \pm 0.11)$.}
	\label{fig:SpecBreakDists}
\end{figure}

The possibility exists that a redshift dependent spectral break at $\sim \negthickspace 1\,$TeV could be the result of an effect intrinsic to the observing instrument.  Perhaps the highest energy spectral bins contain a pileup of events that systematically hardens the spectrum.  We have investigated this issue by examining two distributions.  The first distribution of relevance is the spectral break (as shown in Figure \ref{fig:TeVBreakVsZ}) as a function of the spectral index obtained when fitting the blazars listed in Table \ref{tab:BlazarSample} with a single power-law.  This is plotted in Figure \ref{subfig:SpecBreakVsGamma}.  A pileup effect would be most pronounced for soft sources due to fewer statistics at the highest energies.  Consequently, if such an instrumental pileup were present, one would expect to see an increase in the measured spectral break as the spectral index increases (softens).  The best linear fit to the data shown in \ref{subfig:SpecBreakVsGamma} is given by $\Delta\Gamma(\Gamma) = (1.28 \pm 0.68)\Gamma - (3.15 \pm 2.15)$, with a fit probability of 22\%.  This suggests that any pileup as a function of spectral index is approximately a $2 \sigma$ effect.

The next important thing to establish is whether or not this effect could be responsible for the $\sim \negthickspace 3\sigma$ trend of increasing spectral break with redshift seen in Figure \ref{fig:TeVBreakVsZ}.  This can be determined by looking at the single power-law spectral index distribution as a function of redshift (Figure \ref{subfig:GammaVsZ}).  In order for a systematic pileup effect to be responsible for the spectral break versus redshift distribution observed, the spectral indices of blazars would have to be increasingly soft as the source redshift increases.  As can be seen in Figure \ref{subfig:GammaVsZ}, this is not the case.  The best linear fit to the spectral index versus redshift distribution is given by $\Gamma(z) = (0.46 \pm 0.87)z + (3.05 \pm 0.11)$  and is clearly consistent with being flat.  This may seem surprising given that EBL absorption is expected to result in softer spectra as source redshifts increase.  However, hard spectrum blazars are more readily detected by IACTs given that they generally have higher fluxes than their soft spectrum counterparts at very high energies.  This results in the selection effect that blazars of higher redshift are more likely to be intrinsically hard.

One can conclude from these arguments that a pileup effect does not systematically produce a redshift dependent spectral break in IACT spectra.  Furthermore, the spectra used in Method 2 come from three different IACTs all of which use different energy and spectral reconstruction techniques.  This further reduces the likelihood that a redshift dependent spectral break could be purely an instrumental effect.

\end{document}